\documentclass[prl,twocolumn]{revtex4-1}
\pdfoutput=1

\usepackage{amsmath,amssymb,amscd}         
\usepackage{dsfont}

\usepackage{hyperref}
\hypersetup{
    colorlinks=true,
    citecolor=blue,
    linkcolor=blue,
    urlcolor=magenta,
}
\usepackage[pdftex]{graphicx}

\usepackage{tikz}
\tikzset{every picture/.style={line width=0.75pt}} 
\usepackage{tikz-cd, tikz-feynman} 
\tikzfeynmanset{warn luatex=false}

\newcommand{\bfx}{\mathbf{x}}

\newcommand{\bfk}{\mathbf{k}}

\newcommand{\bfq}{\mathbf{q}}
\newcommand{\bfK}{\mathbf{K}}

\newcommand{\SBD}{\mathcal{S}} 
\newcommand{\SUD}{\mathtt{S}}  
\newcommand{\fUD}{\mathtt{f}}  
\newcommand{\m}{\mu}  
\newcommand{\G}{\mathcal{G}} 
\newcommand{\GUD}{\mathtt{G}}  

\begin{document}

\title{A de Sitter $S$-matrix for the masses}


\author{Scott Melville}
\affiliation{Queen Mary University of London, Mile End Road, London, E1 4NS, U.K.}

\author{Guilherme L. Pimentel}
\affiliation{Scuola Normale Superiore and INFN, Piazza dei Cavalieri 7, Pisa, 56126, Italy}

\date{September 2023}

\begin{abstract}
\noindent We define an $S$-matrix for massive scalar fields on a fixed de Sitter spacetime, in the expanding patch co-ordinates relevant for early Universe cosmology.
It enjoys many of the same properties as its Minkowski counterpart, for instance: it is insensitive to total derivatives and field redefinitions in the action; it can be extracted as a particular ``on-shell'' limit of  time-ordered correlation functions; and for low-point scattering, kinematics strongly constrains its possible structures. 
We present explicit formulae relating the usual observables---in-in equal-time correlators and wavefunction coefficients at the conformal boundary---to $S$-matrix elements. 
Finally, we discuss some of the subtleties in extending this $S$-matrix to light fields (in the complementary series).  
\end{abstract}

\maketitle 


\noindent The best understood observables in quantum field theory are asymptotic. 
As separations between detectors are taken to be infinitely large in a controllable way, the overlaps between states can be computed at weak coupling using perturbation theory. 
This information is contained in the $S$-matrix of the theory. 
Moreover, as gravity forbids the existence of local operators, asymptotic observables might be the only ones that ultimately make sense in a theory of gravity. 
While the $S$-matrix is well understood in Minkowski spacetime \cite{Eden:1966dnq, Elvang:2015rqa} and, to some extent, in AdS \cite{Giddings:1999qu, Heemskerk:2009pn,Penedones:2010ue,Fitzpatrick:2011hu,Raju:2012zr,Paulos:2016fap}, a de Sitter counterpart has received much less attention (though see \cite{Spradlin:2001nb, Bousso:2004tv, Dvali:2017eba, Cheung:2022pdk}), despite much recent activity in the study of cosmological observables in perturbation theory (for reviews see \cite{Baumann:2022jpr,Benincasa:2022gtd}).  

In this paper we present a concrete definition of the $S$-matrix in de Sitter spacetime that is directly applicable to primordial cosmology.
We work throughout in the expanding Poincar\'e patch, using co-ordinates $ds^2 = \tau^{-2} \left( -d \tau^2 + d \bfx^2 \right)$ where the conformal time $\tau < 0$ (in units where the Hubble rate $H=1$). 
Our approach closely parallels the standard treatment of the Minkowski $S$-matrix, and is in the same spirit as the pioneering work of Marolf, Morrison and Srednicki for de Sitter in global slicing \hfill \cite{Marolf:2012kh}. 
\hfill Our \hfill $S$-matrix \hfill elements \hfill are \hfill insensitive \hfill to \\ total derivatives and field redefinitions in the Lagrangian, with a simple crossing relation between different channels.
An advantage of working in the expanding Poincar\'{e} patch is that our $S$-matrix elements explicitly connect to the inflationary wavefunction and primordial non-Gaussianities that characterise the early Universe. 

This sheds light on previously observed obstacles to defining an $S$-matrix for light fields on de Sitter.
For particular mass values, the wavefunction develops divergences at late times that require holographic renormalisation \cite{Bzowski:2015pba,Bzowski:2018fql}. 
The relation we derive between the wavefunction and $S$-matrix makes it clear that similar divergences must appear in the $S$-matrix elements for light fields with masses $m < d/2$ in $d$ spatial dimensions. 

Our first goal is therefore to define and study the finite $S$-matrices that describe the scattering of {\it massive scalars} on a fixed de Sitter background---specifically, scalars with masses $m \geq d/2$ and therefore in the principal series of irreducible representations of the de Sitter group. 
We will then return to the issues posed by light scalars and show that certain $S$-matrix elements remain finite for specific interactions.
Although we study only scalar fields here, we believe that the inclusion of spin is a technical hurdle that we can overcome, at least for massive particles. 
The case of dynamical gravity is more subtle---see \cite{Dvali:2020etd} for arguments that the $S$-matrix might not even exist in that case. 
While these extensions are interesting and deserve further investigation, our results 
provide a first step towards understanding cosmological observables through their underlying $S$-matrix description.

\section{Defining an $S$-matrix}

\noindent We will begin with an abstract discussion of what the different $S$-matrix elements represent, and then provide a concrete definition in terms of field theory correlators. \\

\noindent {\it Choice of basis} --- In order to define $S$-matrix elements, one requires a basis of ``in'' and ``out'' states. 
On Minkowski spacetime, there is a natural choice: using the particle eigenstates $|n \rangle$ of the free theory (i.e. eigenstates of the free Hamtilonian), we can define the ``in''/``out'' states of the interacting theory to be those which coincide with $|n\rangle$ in the far past/future. 
However, on de Sitter the number of particles is not conserved due to gravitational particle production: the state $| n , \tau_* \rangle$ which contains $n$ particles at time $\tau_*$ is \emph{not} an eigenstate of the free Hamiltonian at later times $\tau \neq \tau_*$, since $n$ particles will generally evolve into a superposition of more/fewer particles due to the expansion of spacetime.
This presents a choice in how we define our asymptotic states.

One natural choice is to define the ``in''/``out'' states of the interacting theory to be those that coincide with $| n , -\infty \rangle$ in the far past/future.
We denote the resulting $S$-matrix elements by 
\begin{align}
\SBD_{n' \to n}  \equiv  {}_{\rm out} \langle n , - \infty | n' , -\infty  \rangle_{\rm in} \; .
\label{eqn:SBD_def}
\end{align}
Equivalently, these matrix elements are the coefficients in the expansion
\begin{align}
 | n' , -\infty  \rangle_{\rm in} = \sum_{n}  \SBD_{n' \to n}  \, | n , -\infty \rangle_{\rm out} \; , 
\label{eqn:Sdef2}
\end{align}
where the sum over $n'$ includes integrals over all momenta and other quantum numbers of the $n$ particles. 
We will therefore refer to this $\SBD$ as the {\bf Bunch-Davies $S$-matrix}, since it describes the time evolution of the Bunch-Davies vacuum state $| 0 , -\infty \rangle$ (and its excitations) in the interacting theory \footnote{
Note that we work throughout in the Heisenberg picture. In the Schr{\"{o}}dinger picture, \eqref{eqn:Sdef2} corresponds to 
expanding $\hat{U} ( 0, - \infty) \hat{a}_{n'}^\dagger ... \hat{a}_{1'}^\dagger | \Omega \rangle$ in terms of the states $\hat{U}_{\rm free} (0 , - \infty ) \hat{a}_n^\dagger ... \hat{a}_1^\dagger | \Omega \rangle$, where $|\Omega \rangle$ is the Bunch-Davies vacuum state in the far past and $\hat{U} ( \tau_2, \tau_1)$ is the time evolution operator from $\tau_1$ to $\tau_2$. 
}.  

Another, equally natural, choice is to instead define the ``out'' states of the interacting theory as those that coincide with $| n , 0\rangle$ in the far future.
This produces a different set of $S$-matrix elements,
\begin{align}
 \SUD_{n' \to n}  \equiv {}_{\rm out} \langle n , 0 | n' , -\infty  \rangle_{\rm in} \; .
\label{eqn:SUD_def}
\end{align}
We will refer to this $\SUD$ as the {\bf Unruh-DeWitt $S$-matrix}, since it describes scattering from a state containing $n'$ particles to a state containing $n$ particles, as measured by an Unruh-DeWitt detector in the far past/future.

These two sets of $S$-matrix elements are ultimately related by a Bogoliubov transformation (which maps $|n , - \infty \rangle$ to $|n , 0\rangle$ in the free theory). 
We will initially focus on the Bunch-Davies $S$-matrix because it has:
\begin{itemize}
 
 \item[(i)] no particle production in the free theory, i.e. without interactions, all off-diagonal $\SBD_{n' \to n}$ vanish by construction, 
 
 \item[(ii)] a simple crossing relation that exchanges particles between the in- and out-states, 
 
 \item[(iii)] a closer connection to the wavefunction and in-in correlators used in inflationary cosmology. \\
 
\end{itemize}

\noindent {\it The $S$-matrix from a reduction formula} --- 
The $S$-matrix overlap in \eqref{eqn:SBD_def} can be extracted from time-ordered correlation functions by ``amputating'' the external legs and going ``on-shell,'' in analogy with the LSZ formula in flat space. 
Concretely, consider a real scalar field $\phi (\tau, \bfx)$ of mass $m^2 = (d/2)^2 + \m^2$.
We split the action $S = S_{\rm free} + S_{\rm int}$, where the free quadratic action is
\begin{align}
S_{\rm free}  = -\int d\tau d^d \bfx \; \sqrt{-g} \; \tfrac{Z^2}{2} \left( g^{\alpha \beta} \partial_\alpha \phi \partial_\beta \phi + m^2 \phi^2 \right)  
\label{eqn:Sfree}
\end{align}
and $S_{\rm int}$ contains all non-linear interactions.
Canonical quantisation then proceeds as usual: we first quantise the free theory $S_{\rm free}$, which can be done exactly, and then include the effects of $S_{\rm int}$ as a small perturbation. 

Performing a Fourier transform from position $\bfx$ to momentum $\bfk$, the free equation of motion for the canonically normalised $\varphi (\tau ,\bfk ) \equiv (- \tau )^{-d/2} \phi ( \tau, \bfk )$ is \footnote{
The Fourier transform is performed using the flat metric, so $\phi (\tau, \bfk) = \int d^d \bfx \, e^{  i \bfk \cdot \bfx} \phi (\tau ,  \bfx )$ where $\bfk \cdot \bfx = k^i \delta_{ij} x^i$ is $\tau$-independent. 
We also abuse notation and use the same symbol to denote functions in different representations: for instance $\phi (\tau, \bfx)$ and $\phi ( \tau, \bfk)$ are of course different functions (one is the field in position space, the other is the field in momentum space). 
}
\begin{align}
	\mathcal{E} [ k \tau ] \, \varphi ( \tau , \bfk )  \equiv  \left[ 
	\left( \tau \partial_\tau \right)^2 + k^2 \tau^2 + \m^2 \right] \varphi ( \tau , \bfk ) = 0 \; . 
 \label{eqn:eom_def}
\end{align}
In the Heisenberg picture, the time evolution of the $\hat{\varphi}$ operator can therefore be written as
\begin{align}
\hat{\varphi} ( \tau , \bfk )=  f^- ( k \tau ) \hat{a}_{ - \bfk}  + f^{+} ( k \tau  ) \hat{a}_{\bfk}^{ \dagger} \; , 
\end{align}
where the mode functions satisfy the free equation of motion $\mathcal{E} [ k \tau ] f^{\pm} ( k \tau ) = 0$ with the boundary condition
\begin{align}
0 = \left(  \tau \partial_\tau  \pm i \sqrt{k^2 \tau^2 + \m^2 } \right)  f^{\pm}  ( k \tau ) |_{\tau = \tau_*}  \; , 
\label{eqn:vac_cond}
\end{align}
which ensures that $\hat{a}_{\bfk}$ diagonalises the free Hamiltonian at time $\tau_*$. 
Consequently, $\hat{a}_{\bfk}  | 0 , \tau_* \rangle  =  0$ defines the instantaneous vacuum state $| 0 , \tau_* \rangle$ (the state with the lowest energy at time $\tau_*$), and $\hat{a}^\dagger_{\bfk}$ creates a ``particle'' of momentum $\bfk$ at time $\tau_*$ (an excited eigenstate of the Hamiltonian at time $\tau_*$). 
A complete basis of states for the Hilbert space is then provided by
\begin{align}
	| n , \tau_*  \rangle \equiv \hat{a}^\dagger_{n} ... \hat{a}^\dagger_{1} | 0  \rangle \; ,   
\end{align}
where the label on each $\hat{a}^\dagger$ denotes both the momenta and all other quantum numbers (e.g. mass) of that particle, and $| n \rangle$ denotes the complete list of this $n$-particle data. 
For the Bunch-Davies $S$-matrix, we impose the vacuum condition at $\tau_* \to - \infty$, which corresponds to Hankel mode functions,
\begin{align}
f^+ ( z  ) &\equiv  \frac{  \sqrt{\pi} }{2 i Z } e^{+  \frac{\pi}{2}   \m }  H_{i \m}^{(2)} \left( - z \right) = \left[ f^- ( z^* ) \right]^* \; , 
\label{eqn:f1_def}
\end{align}
which have been normalised so that \footnote{
The real constant $Z$ describes the power spectrum at early times,
\begin{align}
\lim_{\tau \to -\infty} \left \langle  \hat{\varphi} \left(\tau,  \bfk' \right) \hat{\varphi} \left( \tau , \bfk \right) \right \rangle  = \frac{1}{2k Z^2} \, \delta^d \left( \bfk + \bfk' \right) \; , 
\end{align}
and is fixed by the overall normalisation of the $\phi$ kinetic term in the Lagrangian. 
The phase of $f^{\pm}$ has been fixed so that the crossing relation \eqref{eqn:f_cross} has a trivial phase. 
}
\begin{align}
i Z^2 f^- \left( k \tau  \right) ( \overset{\leftrightarrow}{ \tau \partial_\tau} ) \hat{\varphi} ( \tau, \bfk )  =  \hat{a}_{\bfk}^\dagger \; ,
 \label{eqn:fvarphi_to_a}
\end{align}
where $[ \hat{a}_{\bfk'} , \hat{a}^\dagger_{\bfk} ] = (2 \pi )^d \delta^d \left( \bfk + \bfk' \right)$. 

In the interacting theory, we now seek to define states $| n , -\infty  \rangle_{\rm in}$ and $| n , -\infty  \rangle_{\rm out}$, which coincide with the state $| n , -\infty  \rangle$ as $\tau \to -\infty$ and $\tau \to 0$ respectively.
By ``coincide,'' we mean for instance that 
\begin{align}
\lim_{\tau \to -\infty}  \langle \alpha | \hat{\mathcal{O}} (\tau) |  0 , - \infty \rangle_{\rm in} 
=
\lim_{\tau \to - \infty} \langle \alpha | \hat{\mathcal{O}} ( \tau) | 0 , - \infty  \rangle 
\end{align}
for any operator $\hat{\mathcal{O}}$ and normalisable state $| \alpha \rangle$ in the Heisenberg picture (and strictly speaking the limit on the right-hand-side should be $\tau \to - \infty (1 - i \epsilon )$ to ensure convergence). 

For brevity, from now on we will denote Bunch-Davies asymptotic states as $| n \rangle_{\rm in}$ and $|n  \rangle_{\rm out}$.

The idea is then to find an operator which acts on $| 0 \rangle_{\rm in}$ to create the 1-particle in-state $| 1  \rangle_{\rm in}$. 
We claim that
\begin{align}
 \langle \alpha | 1  \rangle_{\rm in} 
= \lim_{\tau \to - \infty } \langle \alpha |  \, i Z^2 f^- ( k \tau) \overset{\leftrightarrow}{ ( \tau \partial_{\tau} ) } \hat{\varphi}(\tau , \bfk) \,  | 0  \rangle_{\rm in}  \;  	
\label{eqn:LSZin}
\end{align}
for any normalisable state $| \alpha \rangle$. 
Clearly the right-hand-side generates a 1-particle state in the free theory thanks to \eqref{eqn:fvarphi_to_a}, and we argue in the Appendix that in the limit $\tau \to - \infty$ the interactions turn off sufficiently quickly that this operator produces the desired 1-particle in-state. This is the same ``adiabatic hypothesis" used to define LSZ operators on Minkowski.
Then any $| n  \rangle_{\rm in}$ can be constructed by repeated application of \eqref{eqn:LSZin}. 

Similarly, we claim that for the out-states,
\begin{align}
- {}_{\rm out} \langle 1 | \alpha  \rangle 
	= \lim_{\tau \to 0 } {}_{\rm out} \langle 0 | i Z^2 f^+ ( k \tau) \overset{\leftrightarrow}{ ( \tau \partial_{\tau} ) } \hat{\varphi}^\dagger (\tau , \bfk) | \alpha  \rangle  \;  	
	\label{eqn:LSZout}
\end{align}
for any normalisable state $| \alpha \rangle$. Again this requires that the interactions turn off sufficiently quickly at late times, which is the case for massive fields in the principal series and derivatively coupled fields in the complementary series (see the Appendix for details).
We also have the useful corollary that these operators can be used to annihilate ${}_{\rm out} \langle 0 |$ and $| 0 \rangle_{\rm in}$, for instance:
\begin{align}
 \lim_{\tau \to 0 } {}_{\rm out} \langle  0  |  f^- ( k \tau) \overset{\leftrightarrow}{ ( \tau \partial_{\tau} ) } \hat{\varphi} (\tau , \bfk) | \alpha \rangle &= 0 \; .   \label{eqn:LSZ_coro_1} 
\end{align}

To relate the $S$-matrix elements to a field correlator, we can now follow the analogous steps as in flat space. 
By applying \eqref{eqn:LSZin}, we see that any particle from the in-state can be replaced by a field insertion, 
\begin{align}
&i Z^{-2} \; {}_{\rm out} \langle n'  | n  \rangle_{\rm in}
\label{eqn:LSZ_derivation} \\
&= - \lim_{\tau \to - \infty} {}_{\rm out} \langle n' | \; f^- ( k_n \tau) \overset{\leftrightarrow}{ ( \tau \partial_{\tau} ) } \hat{\varphi} (\tau , \bfk_n ) \; |  n-1  \rangle_{\rm in}  \nonumber  \\
 &=  \int_{- \infty}^{ 0 } d \tau \; \partial_\tau \left[ {}_{\rm out} \langle n'  | \; f^- ( k_n \tau) \overset{\leftrightarrow}{ ( \tau \partial_{\tau} ) } \hat{\varphi} (\tau , \bfk_n) \;  | n-1  \rangle_{\rm in}   \right]    \nonumber \\ 
 &=  \int_{-\infty}^0 \frac{d \tau}{\tau} \;\;   f^- ( k_n \tau  )\,  \hat{\mathcal{E}} [ k_n \tau ]   \; {}_{\rm out} \langle n'  | \;  \hat{\varphi} (\tau, \bfk_n) \;   |  n-1  \rangle_{\rm in}  . \nonumber  
\end{align}
Note that in going to the penultimate line we have assumed that none of the momenta in $\langle n' |$ coincide with those in $\bfk_n$, and therefore we can use \eqref{eqn:LSZ_coro_1} to discard the $\tau \to 0$ limit of the integral. This amounts to considering the \emph{connected} part of the $S$-matrix element \footnote{
One exception is the trivial $1 \to 1$ scattering amplitude, for which translation invariance requires that $\bfk = \bfk'$ and an additional boundary term must be included in \eqref{eqn:LSZ_derivation}---see the Appendix for more details about these disconnected boundary contributions.
}. 
Proceeding in the same way for each particle in $|n \rangle$ and $\langle  n' |$, one can reduce the right-hand side to the vacuum expectation of a (time-ordered) product of field insertions. 
We therefore define the correlator, the amputated correlator, and the connected part of the Bunch-Davies $S$-matrix element by 
\begin{widetext}
\begin{align}
G_{n' \to n} 
&\equiv {}_{\rm out} \langle 0  | \; T\, 
 \prod_{b=1}^{n} \hat{\varphi}^\dagger  (\tau_{b}, \bfk_{b})  \prod_{b'=1}^{n'}   \hat{\varphi} (\tau'_{b'} , \bfk'_{b'} ) \;  |  0  \rangle_{\rm in}  \; , \nonumber \\ 
\G_{n' \to n} &\equiv  \left[ \prod_{b=1}^{n}  i Z^2 \hat{\mathcal{E}} [ k_b \tau_b] \right] \left[ \prod_{b'=1}^{n'}  i Z^2 \hat{\mathcal{E}} [ k_{b'}' \tau_{b'}' ] \right] G_{n \to n'} \; ,  \label{eqn:LSZ} \\
\SBD_{n' \to n} &= 
 \left[ \prod_{b=1}^n  \int_{-\infty}^0 \frac{d \tau_b}{-\tau_b} \;  f^+ ( k_b \tau_b  )  \right] 
\left[ \prod_{b'=1}^{n'} \int_{-\infty}^0 \frac{d \tau_b'}{-\tau_b'} \; f^- ( k_b' \tau_b'  )  \right] 
  \, \G_{n' \to n} \nonumber
\end{align}
\end{widetext}
where $T$ represents time-ordering in $\tau$, and the lower limits of the time integrals are understood to be $\tau \to -\infty (1 \mp i \epsilon)$ for the in/out-going particles.

Formula \eqref{eqn:LSZ} is our prescription for the de Sitter $S$-matrix: in words, one should first compute the time-ordered correlation function, then apply the classical equations of motion to each field (this ``amputates'' its external leg from any Feynman diagram) and finally perform an integral transform using Hankel mode functions (this puts the external legs ``on-shell''). \\ 

\noindent {\it Perturbation theory} --- To compute $\SBD_{n' \to n}$ in perturbation theory, one can go to the interaction picture and expand in Feynman diagrams in which:
\begin{itemize}

\item outgoing external lines represent the free mode function $f^+ ( k\tau )$, 

\item ingoing external lines represent the free mode function $f^- ( k\tau )$,

\item internal lines represent the free theory propagator
\begin{align}
 \langle 0  | T \, \hat{\varphi} (\tau , \bfk ) \hat{\varphi} (\tau' , \bfk' ) | 0  \rangle 
 \equiv
 G_{2} \left( 
 k \tau , k \tau' \right) (2 \pi )^d \delta^d \left( \bfk + \bfk' \right)  , 
\nonumber 
\end{align}

\item $n$-point vertices represent local interactions involving $n$ powers of $\varphi$, and multiply the above propagators by a vertex factor of $i\, \delta^n S_{\rm int}/\delta \varphi^n$,  

\item finally, all internal times and momenta are integrated over. 

\end{itemize}
Regardless of the contention about stability/existence of de Sitter spacetime in a quantum theory of gravity, this $\SBD_{n' \to n}$ certainly exists perturbatively. 
For instance, a local interaction $\sqrt{-g} \tfrac{\lambda_n}{n!} \phi^n$ in $S_{\rm int}$ will produce a calculable ``contact'' contribution to $\SBD_{0 \to n}$ of
\begin{align}
    \SBD^{\rm cont}_{0 \to n} =  i \lambda_n  \int_{-\infty}^0 \frac{d \tau}{ - \tau} ( - \tau )^{\frac{d}{2} (n-2)} \prod_{b=1}^n f^+ ( k_b \tau_b ) \; , 
    \label{eqn:Sn_cont}
\end{align}
where we have suppressed the overall momentum-conserving $\delta$-function. 
It will also produce ``exchange" diagram contributions to higher-order $S$-matrix elements: for instance, a $\phi^3$ interaction will give the following contribution to $\SBD_{0 \to 4}$:
\begin{widetext}
\begin{align}
\SBD_{0 \to 4}^{\rm exch}  =  - \lambda_3^2 \int_{-\infty}^0 \frac{d \tau}{-\tau}  \, ( - \tau )^{d/2} \, \int_{-\infty}^{0} \frac{d \tau'}{- \tau'} ( - \tau' )^{d/2} \,  f^+ (k_1 \tau) f^+ (k_2 \tau) G_2 \left( k_s \tau, k_s \tau' \right)  f^+ ( k_3 \tau') f^+ ( k_4 \tau') + 2 \, \text{perm.} \label{eqn:S4_exch}
\end{align}
\end{widetext}
where $k_{s} \equiv |\bfk_1 + \bfk_2|$ and ``2 perm.'' denotes the $t$ and $u$-channel contributions (again omitting a $\delta$-function).  \\

\noindent {\it The other $S$-matrix} --- To extract the connected part of the Unruh-DeWitt $S$-matrix \eqref{eqn:SUD_def} from a field correlator, we must make two changes to the LSZ formula \eqref{eqn:LSZ}: 
\begin{itemize}

\item[(i)] the mode functions for the in- and out-going particles should be replaced by
\begin{align}
    f^- ( k \tau ) &\to \tfrac{1}{\alpha}  \fUD^- ( k \tau ) \; , &f^+ ( k \tau ) &\to \tfrac{1}{\alpha} f^+ ( k\tau )  \; , 
    \label{eqn:UD_replacement}
\end{align}
where $\fUD^{\pm}$ solves the free equation of motion with the vacuum condition \eqref{eqn:vac_cond} imposed at $\tau_* = 0$, since the operators $\fUD^- \overset{\leftrightarrow}{ ( \tau \partial_{\tau} ) } \hat{\varphi}$ and $f^+ \overset{\leftrightarrow}{ ( \tau \partial_{\tau} ) } \hat{\varphi}$ annihilate ${}_{\rm out} \langle 0 , 0|$ and $|0,-\infty \rangle_{\rm in}$ respectively. Concretely, $\fUD^{-}$ is a Bessel function,
\begin{align}
 \fUD^{\pm} ( k \tau ) = \frac{ \sqrt{\pi} }{ Z \sqrt{ 2 \sinh ( \m \pi )} }\,  J_{\mp i \m} ( - k \tau ) \; , 
\end{align}
and is related to the previous Hankel mode function by the Bogoliubov transformation
\begin{align}
 f^+ ( k \tau  ) = \alpha \; \fUD^+ ( k \tau ) + \beta \; \fUD^- (k \tau ) \; , 
 \label{eqn:Bogoliubov}
\end{align}
where $|\alpha|^2 - | \beta|^2 = 1$ \footnote{
The Bogoliubov coefficients are given explicitly by $\alpha = e^{+ \mu \pi/2} / \sqrt{2 \sinh ( \mu \pi ) }$ and $\beta = e^{- \mu \pi/2} / \sqrt{2 \sinh ( \mu \pi ) }$
}. 
The factors of $\alpha$ in \eqref{eqn:UD_replacement} arise from writing $f^- \overset{\leftrightarrow}{ ( \tau \partial_{\tau} ) } \hat{\varphi} | n \rangle_{\rm in} = \frac{1}{\alpha} \fUD^- \overset{\leftrightarrow}{ ( \tau \partial_{\tau} ) } \hat{\varphi} | n \rangle_{\rm in}$ in the first step of the LSZ reduction \eqref{eqn:LSZ_derivation}, where again we focus on the connected component (see \eqref{eqn:disc_comp} for the disconnected contributions). 

\item[(ii)] the ${}_{\rm out} \langle 0 , -\infty|$ bra in the time-ordered correlator $G_{n' \to n}$ should be replaced by ${}_{\rm out} \langle 0 , 0 |$, which changes the boundary condition for internal lines.
Concretely, the propagator for the Bunch-Davies $S$-matrix can be written in terms of the Hankel mode functions as
\begin{align}
 G_2 ( k \tau_1 , k \tau_2 ) = f^- ( k \tau_{>} ) f^+ ( k \tau_{<} ) \; ,
\end{align}
while for the Unruh-DeWitt $S$-matrix one must instead use the propagator 
\begin{align}
 \GUD_2 ( k \tau_1 , k \tau_2 ) = \tfrac{1}{\alpha} \, \fUD^- ( k \tau_{>} ) f^+ ( k \tau_{<} )  \; , 
\end{align}
where $\tau_{> (<)}$ is the greater (lesser) of $\tau_1$ and $\tau_2$. 

\end{itemize}

\section{Properties of the $S$-matrix}

\noindent Before giving explicit examples of these $S$-matrix elements, let us list some model-independent properties. \\

\noindent {\it Particle production} ---
Already in the free theory, there is an important difference between the Bunch-Davies and Unruh-DeWitt $S$-matrix elements. 
The only non-zero Bunch-Davies $S$-matrix with two particles is
\begin{align}
\SBD_{1 \to 1} &= (2 \pi )^d \delta^d ( \bfk - \bfk' ) \; ,
\label{eqn:S1to1}
\end{align}
and simply reflects the normalisation we have chosen for the asymptotic states, namely that $[ \hat{a}_{\bfk'} , \hat{a}^{\dagger}_{\bfk} ] = \delta^d ( \bfk + \bfk' ) $. 
On the other hand, the Unruh-DeWitt $S$-matrix has a non-zero $\SUD_{0 \to 2}$ since the states $| 0 , - \infty \rangle$ and $\langle 2 , 0 |$ are not orthogonal. In fact, given the Bogoliubov transformation \eqref{eqn:Bogoliubov}, their overlap is
\begin{align}
\SUD_{0 \to 2}
 &=  \frac{\beta}{\alpha} \, (2 \pi )^d \delta^d \left( \bfk_1 + \bfk_2 \right)  \; . 
\end{align}
The Bogoliubov coefficient $\beta$ therefore characterises the rate of particle production in the free theory, as measured by an Unruh-DeWitt detector (and $\beta/\alpha = e^{- \mu \pi}$ is the characteristic Boltzmann factor which suppresses the production of heavy states).
One special feature of the Bunch-Davies basis for the $S$-matrix is that this particle production is accounted for by the choice of asymptotic states: the elements $\SBD_{n' \to n}$ are the probability that an initial $n'$-particle state will scatter into the jets of multi-particle ``stuff'' which would have been created by the free propagation of $n$ particles through the expanding spacetime medium.

For $S$-matrix elements with more than two particles, this free theory particle production shows up as additional contributions to the disconnected parts of the Unruh-DeWitt $S$-matrix: see Figures~\ref{fig:free} and ~\ref{fig:int} in the Appendix for a concrete example. 
In the interacting theory, there is additional particle production due to the interactions in $S_{\rm int}$. These appear explicitly in both the Bunch-Davies and the Unruh-DeWitt bases: for instance both $\SBD_{0 \to n}$ and $\SUD_{0 \to n}$ are generically non-zero when $S_{\rm int}$ contains $n$-point interactions. \\

\noindent {\it Antipodal singularities} --- Performing the inverse Fourier transform from each momentum $\bfk$ back to a position $\bfx$, both the Bunch-Davies propagator $ ( \tau_1 \tau_2 )^{d/2} G_2 ( k \tau_1, k \tau_2)$ and the Unruh-Dewitt propagator $ ( \tau_1 \tau_2 )^{d/2} \GUD_2 ( k \tau_1, k \tau_2)$  become functions of the invariant chordal separation between $( \tau_1 , \bfx_1 )$ and $(\tau_2, \bfx_2 )$,
\begin{align}
 \cosh \, \sigma  \equiv 1 + \frac{ ( \tau_1 - \tau_2 )^2  - | \bfx_1 - \bfx_2 |^2  }{2 \tau_1 \tau_2} \; . 
 \label{eqn:sigma_def}
\end{align}
Explicitly, these functions can be written in terms of the associated Legendre functions \footnote{
Our conventions for the Legendre functions are 
\begin{align*}
P^n_{i \mu - \tfrac{1}{2}} (z) 
 &= \tfrac{1}{\Gamma \left( 1 - n \right) }
  \left( \tfrac{z + 1}{ z - 1 } \right)^{n/2}  {}_2 F_1 \left( \tfrac{1}{2} - i \mu , \tfrac{1}{2} + i \mu ;	1- n ; \tfrac{1 - z}{2} \right)  \\
  \tfrac{ Q^n_{i \mu - \tfrac{1}{2}} (z) }{ \left( z^2 - 1 \right)^{n/2} } 
&= \tfrac{ e^{i \pi n} }{ 2^{n+1} }  \tfrac{ \Gamma \left(i \mu
   +n+\frac{1}{2}\right) \Gamma \left( i \mu + \frac{1}{2} \right) }{ \Gamma \left( 2 i \mu + 1 \right) }  \left(  \tfrac{ z - 1 }{2} \right)^{- i \mu - \frac{1}{2} - n}   \\
 &\quad\times   {}_2 F_1 \left( i \mu + n + \tfrac{1}{2} , i \mu + \tfrac{1}{2} , 2 i \mu + 1 ; \tfrac{2}{1 - z } \right)  \; .
 \end{align*}
 \\[-12pt] \!
}
\begin{align}
\GUD_2 ( \cosh \sigma ) &= \frac{ Q^{\frac{d}{2} -\frac{1}{2}}_{i \mu - \frac{1}{2}} ( \cosh \sigma )  }{ ( 2 \pi )^{\frac{d+1}{2}}  ( \sinh \sigma )^{\frac{d}{2} - \frac{1}{2}} } \; ,  \nonumber \\
G_2 ( \cosh \sigma ) &= \frac{ \pi }{ 2 \cosh ( \pi \mu ) } \frac{ P^{\frac{d}{2} -\frac{1}{2}}_{i \mu - \frac{1}{2}} ( - \cosh \sigma  )  }{ ( 2 \pi )^{\frac{d+1}{2}} ( \sinh \sigma )^{\frac{d}{2} - \frac{1}{2}} } \; , 
\end{align}
which are often expressed in terms of either Gegenbauer functions or hypergeometric ${}_2 F_1$ functions \cite{Chernikov:1968zm, Bunch:1978yq, Allen:1985ux, persee.fr:barb_0001-4141_1968_num_54_1_62238, Polyakov:2007mm}. 

$\cosh \sigma > 1$ ($<1$) corresponds to the two positions being time-like (space-like) separated. Both propagators have a branch point singularity on the lightcone at $\cosh \sigma = 1$, and the branch cut along $\cosh \sigma > 1$ reflects the ambiguity in ordering time-like separated operators. The time-ordering relevant for our $S$-matrix is the prescription that these functions are evaluated at $\cosh (\sigma) - i \epsilon$ and the branch cut is approached from below \cite{Allen:1985wd, Fukuma:2013mx}. 
This is precisely analogous to the singularity structure of the Feynman propagator on Minkowski. 

One respect in which de Sitter differs qualitatively from Minkowski is the existence of an antipodal map: sending $(\tau_1, \bfx_1)$ to its antipodal position in the de Sitter spacetime corresponds to sending $\cosh \sigma \to - \cosh \sigma$.  
The region $\cosh \sigma < -1$ therefore corresponds to $(\tau_2, \bfx_2)$ being time-like separated from the \emph{antipode} of $(\tau_1, \bfx_1)$. While the Bunch-Davies propagator $G_2$ is perfectly regular at $\cosh \sigma = -1$, the Unruh-DeWitt propagator $\GUD_2$ has an additional branch point singularity there. There is no analogue of this antipodal singularity on Minkowski, and the existence of this additional branch cut in $\GUD_2$ is another important difference between the Bunch-Davies and Unruh-DeWitt boundary conditions. 

When Wick rotated to Euclidean AdS, the choice of out-state becomes the choice of boundary condition at spatial infinity. 
$G_2$ corresponds to the walls of the AdS box being ``transparent'', while $\GUD_2$ corresponds to the walls being ``reflecting'' \cite{Avis:1977yn, Allen:1985wd}. In that language, it is the reflecting boundary condition that leads to an antipodal image of the coincident singularity in the propagator. \\

\noindent {\it Crossing} --- At the level of the time-ordered correlator, the only difference between an ``in-going'' or ``out-going'' field is simply our convention for the sign of its momentum, since $\hat{\varphi}^{\dagger} ( \tau, \bfk) = \hat{\varphi} ( \tau, - \bfk )$ for a real scalar field. 
A physical distinction only arises when we put the fields on-shell: we either do this using $f^+$ or $f^-$, which is the analogue of setting the particle energy $= + \sqrt{k^2 + m^2}$ or $= - \sqrt{k^2 + m^2}$ on Minkowksi.
To relate these, we can make use of the Hankel function identity 
\begin{align}
   \lim_{\epsilon \to 0} f^+ ( - z + i \epsilon ) = f^- ( z ) \;  
    \label{eqn:f_cross}
\end{align}
for real $z > 0$, which is closely related to the invariance of $\phi$ under CPT transformations.
In particular, by replacing each $k$ with a new variable $\tilde{k}$ (which is independent of $\bfk$) in the final step of the LSZ procedure, we naturally arrive at the object
\begin{align}
    \tilde{ \SBD }_{n} ( \{ \tilde{k} \} , \{ \bfk \}  )  \equiv \left[ \prod_{b=1}^{n} \int_{-\infty}^0 \tfrac{d \tau_b}{-\tau_b} \; f^+ ( \tilde{k}_b \tau_b  )  \right] 
 \, \G_{n} \left( \{ \tau \} , \{ \bfk \}  \right) \, .
\label{eqn:Stilde_def}
\end{align}
This function of $\tilde{k}$ contains the $S$-matrix elements for all $n_1 \to n_2$ processes with $n_1 + n_2 = n$, since the transformation
\begin{align}
    ( \tilde{k}_b , \bfk_b ) \to ( - \tilde{k}_b, - \bfk_b )
\end{align}
moves a particle from the out-state to the in-state \footnote{ 
For $S$-matrix elements with only two particles, crossing must be applied carefully since momentum conservation fixes $k_1 = k_2$. 
The naive procedure of ``flipping the sign of $k$'' simply maps $\SBD_{0 \to 2}$ to $\SBD_{2 \to 0}$, both of which are zero for the Bunch-Davies $S$-matrix. 
In terms of $\tilde{\SBD}_2$, the boundary term responsible for the non-zero $\SBD_{1 \to 1}$ element is essentially the Wronskian $ f^+ ( \tilde{k}_1 \tau)( \overset{\leftrightarrow}{ \tau \partial_\tau} ) f^+ ( \tilde{k}_2 \tau )$, which indeed vanishes when both $\tilde{k}$ have the same sign but is non-zero when both $\tilde{k}$ have different signs.   
}. 
For now we restrict our attention to $\tilde{k} = \pm k$ (with an appropriate $i \epsilon$), since these are the values at which $\tilde{\SBD}_n$ coincides with an $\SBD_{n_1 \to n_2}$ element. 
We will return to off-shell extensions of the $S$-matrix in the Future Directions section below. 

Crossing is another important difference between the Bunch-Davies and Unruh-DeWitt $S$-matrices.
The crossing operation that maps a particle from the in- to the out- state in $\SUD$ requires the Bogoliubov transformation \eqref{eqn:Bogoliubov}, and as a result there is no longer a simple function like \eqref{eqn:Stilde_def} that interpolates between different scattering channels for the Unruh-DeWitt $S$-matrix elements. \\

 \noindent {\it de Sitter isometries} --- The $(d+1) (d+2)/2$ isometries of de Sitter spacetime in these co-ordinates are:
 \begin{itemize}
 
 \item $d$ spatial translations, which imply conservation of the total momentum $\bfk$, 
 
 \item $d (d-1)/2$ spatial rotations, which implies a dependence on $\bfk_a \cdot \bfk_b$ only, 
 
 \item $1$ dilation transformation, $(\tau, \bfx ) \to ( \gamma \tau , \gamma \bfx )$, 
which is generated in the momentum domain by
\begin{align}
 D [ \tau, \bfk ] = \bfk \cdot \partial_{\bfk}  - \tau \partial_\tau + d \; , 
\end{align}

\item $d$ ``boosts'', characterised by a parameter $\mathbf{v}$,
\begin{align}
 \tau \to  \gamma \, \tau \;\;\;\; , \;\;\;\; \bfx \to \gamma \left(  \bfx  -  \mathbf{v} x^2\right)     \label{eqn:SCT}  \\
\text{where} \;\;\;\;  \gamma = 1/\left( 1 - 2 \mathbf{v} \cdot \bfx + v^2 x^2  )   \right) \; . \nonumber
\end{align}
This is generated in the momentum domain by
\begin{align}
\bfK [ \tau , \bfk ] &=  \bfK [ \bfk ] -  \bfk \, \tau^2 - 2 \tau \partial_\tau \partial_{\bfk} + 2 d \, \partial_{\bfk}
\end{align}
where $\bfK [ \bfk ] =  2 \bfk \cdot \partial_\bfk \;  \partial_{\bfk}   -  \bfk \;  \partial_{\bfk} \cdot \partial_{\bfk}$ is the usual generator of special conformal transformations \footnote{
Note that when acting on a function of only $k = | \bfk|$, the generator can be written as $\bfK [ \bfk ] + d \partial_{\bfk} = \frac{ \bfk }{k^2}  ( k \partial_k )^2$. This is in line with $\bfK$ being the ``momentum'' associated with translations of the inverted position $\bfx / x^2$, since the special conformal transformation is equivalent to an inversion-translation-inversion. 
}. 

\end{itemize}
Since a scalar field $\phi$ is invariant under dilations and boosts, the Ward identities for correlators of the rescaled $\varphi = (- \tau)^{-d/2} \phi$ are
\begin{align}
	 \sum_{b=1}^n \left( D [ \tau_b, \bfk_b ] - \frac{d}{2} \right)  G_n &= 0 \; , \nonumber \\ 
	  \sum_{b=1}^n \left( \bfK [ \tau_b, \bfk_b ]  -  d \, \partial_{\bfk_b}  \right) \, G_n  &=  0  \; . 
\end{align}
Now applying the LSZ formula, and using the fact that $\mathcal{E}$ represents a quadratic Casimir of the de Sitter algebra and hence commutes with all other generators \footnote{
Given the canonical normalisation of $\varphi$, it is the extended generators $D[\tau,\bfk] - d/2$ and $\bfK [ \tau, \bfk] - d \partial_{\bfk}$ that commute with $\mathcal{E} [ k \tau ]$. 
}, we find that the $S$-matrix for de Sitter invariant interactions is constrained by the Ward identities
\begin{align}
 \sum_{b=1}^n \left( \bfk_b \cdot \partial_{\bfk_b} + \frac{d}{2} \right)  \tilde{\SBD}_n &= 0 \; ,  \nonumber \\ 
 \sum_{b=1}^n \left(  \bfK [ \bfk_b ] + d \partial_{\bfk} + \m^2 \frac{\bfk_b}{k_b^2}  \right)  \tilde{\SBD}_n &= 0 \; . 
\end{align}
For instance, consider the contact contribution~\eqref{eqn:Sn_cont}. 
Applying the above dilation, the integrand shifts by a total derivative which does not contribute to the $S$-matrix and so the corresponding Ward identity is satisfied. 
Applying the above boost, since the mode functions transform as
\begin{align}
\left( \bfK [ \bfk ] + d \, \partial_{\bfk} + \m^2 \frac{\bfk}{k^2}  \right) f^{\pm} ( k \tau )  =  - \bfk \tau^2 f^{\pm} ( k \tau )
\end{align}
the corresponding Ward identity is automatically satisfied thanks to momentum conservation. \\

\noindent {\it Total derivatives and field redefinitions} --- One main advantage of the $S$-matrix formalism is that, unlike the Lagrangian, there is no ambiguity due to field redefinitions and total derivatives. 
For instance, consider the following total derivative
\begin{align}
\mathcal{L}_{\rm td} = \sqrt{-g} \, g^{\alpha \beta} \nabla_\alpha \left(  \phi^2 \nabla_\beta \phi \right) \; . 
\label{eqn:Ltd_eg}
\end{align}
It contributes at tree-level to the $S$-matrix only via the boundary terms of the form
\begin{align}
\int_{-\infty}^0 d \tau \, \partial_\tau \left( 
  f^+ ( k_1 \tau) f^+ ( k_2 \tau) \tau \partial_{\tau} \left[  \tau^{d/2} f^+ ( k_3 \tau) \right] 
  \right)
\end{align}
and both limits separately vanish for principal series fields.
It is easy to see that any total covariant derivative of $\phi$'s will similarly give a vanishing contribution to any $S$-matrix element, simply because $\tau^{d/2} f^{\pm} (k \tau)$ vanishes at both integration boundaries. 
So while total derivatives can contribute to the correlator, their contribution vanishes once we go on-shell. 

To show invariance under field redefinitions, it is useful to consider linear and non-linear redefinitions separately. Linear redefinitions are of the form $\phi' = \gamma \phi$. 
Since this produces a new $S_{\rm free}$ with $Z^2 \to Z^2/\gamma^2$, the normalisation of the mode functions changes in such a way that $f^{\pm} \to \gamma f^{\pm}$. So while the correlator of the new fields is $G_n' = \gamma^n G_n$, it produces the same on-shell $S$-matrix defined in \eqref{eqn:LSZ}.
This is also explicit in the example \eqref{eqn:Sn_cont} given above: since this rescaling produces a new $S_{\rm int}$ with $\lambda_n \to \lambda_n/\gamma^n$, we see that the product of $\lambda_n \times ( f^{\pm} )^n$ is insensitive to linear field redefinitions. 
As an example of a non-linear redefinition, consider $
    \phi \to \phi + \gamma \phi^2 \;  
$ 
applied to the simple Lagrangian $\mathcal{L}_{\rm int} =  \sqrt{-g} \frac{\lambda_3}{3!} \phi^3$. 
This produces a new action
\begin{align}
\mathcal{L} \to \mathcal{L} + \gamma Z^2 \mathcal{L}_{\rm td} + \frac{ \gamma Z^2}{ \tau} ( - \tau )^{\frac{d}{2}} \varphi^2  \mathcal{E} \varphi 
-
\frac{ \gamma \lambda_3}{ 2 \tau} ( - \tau )^{d} \varphi^4 
\end{align}
at leading order in $\gamma$.
The total derivative does not contribute to the $S$-matrix, but the new cubic and quartic interactions give equal and opposite contributions to any $S$-matrix element. For instance,
\begin{widetext}
\begin{align}
\tilde{ \SBD }_{4}^{\rm exch} &\to \tilde{ \SBD }_{4}^{\rm exch} +  4  \gamma \lambda_3 {\textstyle \int_{-\infty}^0} \tfrac{d \tau}{ \tau} (- \tau)^{\frac{d}{2}} {\textstyle \int_{-\infty}^0} \tfrac{d \tau'}{ \tau'} (- \tau' )^{\frac{d}{2}} f^+ ( \tilde{k}_1 \tau) f^+ ( \tilde{k}_2 \tau)  Z^2  \mathcal{E} [k_s \tau] G_2 \left( k_s \tau, k_s \tau' \right)  f^+ ( \tilde{k}_3 \tau') f^+ ( \tilde{k}_4 \tau') + \text{2 perm.}    \nonumber \\[8pt]
\tilde{ \SBD }_{4}^{\rm cont} &\to \tilde{ \SBD }_{4}^{\rm cont} +  12 i \gamma \lambda_3 {\textstyle \int_{-\infty}^0} \tfrac{d \tau}{-\tau} \, (- \tau )^d f^+ ( \tilde{k}_1 \tau) f^+ ( \tilde{k}_2 \tau) f^+ ( \tilde{k}_3 \tau) f^+ ( \tilde{k}_4 \tau )  
\label{eqn:S4_redef}
\end{align}
\end{widetext}
exactly cancel since $Z^2 \mathcal{E} [k_s \tau] G_2 ( k_s \tau, k_s \tau') =  i \tau \delta (\tau - \tau' )$ and hence collapses one of the time integrals in the exchange diagram.
The total $S$-matrix $\tilde{\SBD}_4^{\rm cont} + \tilde{\SBD}_4^{\rm exch}$ is therefore unchanged \footnote{
Here it is important that we set each $\tilde{k} = \pm k$, since otherwise $\mathcal{E} [ k \tau ] f^{+} \left( \tilde{k} \tau \right) \neq 0$ and there would be additional contributions to $\tilde{S}_4^{\rm exch}$, i.e. the off-shell extension of the $S$-matrix need not be invariant under field redefinitions.
}. 
This shows that while the split into ``contact'' and ``exchange'' contributions is ambiguous, the sum is invariant under field redefinitions.

\noindent {\it Unique structures} --- This insensitivity to total derivatives and field redefinitions has the important consequence that the 3-point $S$-matrix is unique (up to crossing).
The argument in perturbation theory is straightforward: since any cubic interaction can be integrated by parts into the form $\phi^2 \Box^n \phi$, an arbitrary derivative interaction will contribute to $\tilde{\SBD}_3$ in the same way as $m^{2n} \phi^3$ for some $n$. 
The contact integral~\eqref{eqn:Sn_cont} (and its three crossing images) from $\phi^3$ are therefore the unique kinematic structures which can appear for three particles in perturbation theory.
To go beyond perturbation theory, notice that these integrals correspond to the four possible solutions to the de Sitter Ward identities \cite{Bzowski:2013sza, Bzowski:2015pba}, and therefore any de Sitter invariant set of interactions must produce an $\tilde{S}_3$ of this form. 
This is the analogue of the well-known result on Minkowski that the on-shell 3-point function is fixed uniquely by the spacetime isometries (to be some, possibly mass-dependent, constant). 

An interesting corollary of this is that there is a unique 4-point exchange structure which describes the interaction of two $\phi$'s via the exchange of a field $\sigma$ (again up to crossing). 
Since the most general cubic vertex is equivalent to a sum of interactions of the form $\phi^2 (\Box - m_\sigma^2 )^n \sigma$ after integration by parts, for any $n \geq 1$ this can be exchanged via a field redefinition of the form $\sigma \to \sigma + ( \Box - m_\sigma^2 )^{n-1} \phi^2$, for a quartic interaction $\phi^2 ( \Box - m_\sigma^2 )^n \phi^2$, which corresponds to the contact invariant in \eqref{eqn:Sn_cont}.
This is the analogue of $1/(m_\sigma^2 - s)$ on Minkowski: any $s$-channel exchange diagram for scalars can always be separated into this unique structure plus contact-type contributions. \\

 \noindent {\it Flat space limit} --- To take the flat space limit, we will temporarily restore factors of the Hubble rate $H$. 
 We will also write the conformal time and mass parameter in terms of a new variable $t$ and $m$ using
\begin{align}
H \tau = - e^{ - H t} \;\; \text{and} \;\; \m = \sqrt{ \frac{m^2}{H^2} - \frac{d^2}{4} } \; . 
\end{align}
In the limit $H \to 0$ at fixed $t, k$ and $m$ (where we do not assume any further hierarchy, so e.g. we treat $k$ and $m$ as comparable), the mode functions become \cite{Boerner:1969ff}
\begin{align}
 f^{\pm} ( k \tau ) = e^{\mp i \alpha_0/H}    \frac{ e^{\pm i \Omega_k t } }{ \sqrt{ 2 i \Omega_k } } \left[ 1 + \mathcal{O} \left(  H  \right) \right] \; , 
 \label{eqn:f_flat}
\end{align}
where we have introduced $\Omega_k = \sqrt{ k^2 + m^2 }$.
Up to an overall phase $\alpha_0$ (which does not affect physical observables), the leading order term in \eqref{eqn:f_flat} coincides with the usual Minkowski mode function \footnote{
Explicitly, $\alpha_0 = \Omega_k - m \, \text{arcsinh} \left( \frac{m}{k} \right)$.
}.  
Also note that \eqref{eqn:f_flat} follows from a saddle point approximation of the Hankel function which requires $ k > 0$. Assuming instead that $k < 0$ produces \eqref{eqn:f_flat} with $\Omega_k = - \sqrt{k^2 + m^2}$.  
The crossing transformation $k \to - k$ therefore implements the usual crossing relation in the Minkowski limit. 

Applied to the reduction formula \eqref{eqn:LSZ}, we find that our de Sitter $S$-matrix coincides in the $H \to 0$ limit with the usual Minkowski $S$-matrix, up to an overall phase and with the state normalisation \eqref{eqn:S1to1}. 
Since the Bogoliubov coefficient $\beta \to 0$ in this limit (i.e. the effects of particle production switch off as $H \to 0$), both the Bunch-Davies and the Unruh-DeWitt $S$-matrix elements have the same Minkowski limit. Physically, this reflects the fact that there is no distinction between the vacua $| 0 , - \infty \rangle$ and $|0 ,0 \rangle$ in the flat space limit (in which the Hamiltonian becomes time-independent and there is a unique vacuum state).  
The same is true of the time-ordered correlation functions: for instance, since the chordal separation \eqref{eqn:sigma_def} becomes $
    \cosh \sigma = 1  - H^2 s^2  + \mathcal{O} \left( H^4 \right)
$
in the flat space limit, where $s^2 \equiv - (t_1 - t_2)^2 + |\bfx_1 - \bfx_2 |^2$ is the Minkowski geodesic distance, 
both propagators have the same limiting behaviour at fixed $\m$, namely \footnote{
\eqref{eqn:flat_G2_coin} also demonstrates that the short-distance singularity at $\cosh \sigma \to 1$ in both de Sitter propagators matches that of Minkowski \cite{Allen:1985wd}, as expected since dS is locally flat.
}
\begin{align}
 \lim_{H \to 0} G_2 ( \cosh \sigma) = \lim_{H \to 0} \GUD_2 ( \cosh \sigma)  = \frac{\Gamma \left( \tfrac{d-1}{2} \right) }{ 4 \pi^{\frac{d+1}{2}} } \left( \tfrac{1}{s^2 + i \epsilon} \right)^{\frac{d-1}{2}} ,
 \label{eqn:flat_G2_coin}
\end{align}
which coincides with the massless Minkowski propagator. The massive propagator is obtained by taking $H \to 0$ at fixed $m$.

\section{Some Examples}

\noindent To illustrate some of these features, we now list some simple $S$-matrix elements in particular models. 

Consider a scalar field $\sigma$ with mass $m^2 = (d^2-1)/4$ (i.e. conformal weight $\Delta = (d-1)/2$). 
This complementary series field has arguably the simplest mode function, since the Hankel function at $i\mu = 1/2$ reduces to a plane wave,
\begin{align}
Z_{\sigma} f^{\pm} ( k \tau )  = \frac{ e^{\pm i k \tau} }{ \sqrt{\mp 2 i k \tau } }  \; \text{ when } i \mu = \frac{1}{2} \; . 
\end{align} 
From the interaction Lagrangian $\mathcal{L}_{\rm int} = \sqrt{-g} \frac{\lambda_n}{n!} \sigma^n$, the $n$-point Bunch-Davies $S$-matrix elements are given by
\begin{align}
\SBD_{0 \to n} = \frac{i \lambda_n}{ Z_\sigma^n  } \frac{\Gamma (  j_n  ) }{ ( i k_T )^{ j_n } }   \frac{ ( 2\pi )^d \, \delta^d \left( \sum_{b=1}^4 \bfk_b \right) }{ \prod_{c=1}^n \sqrt{2 i k_c} }
\label{eqn:eg_1}
\end{align}
and its various crossing images, where $k_T = \sum_{b=1}^n k_b$ is the ``total energy'' flowing into this vertex, and the power $j_n \equiv n \left( \frac{d-1}{2} \right) - d$ uniquely satisfies the dilation Ward identity. 
Note that this result is formally infinite whenever the total conformal weight $n \left( \frac{d-1}{2} \right)$ coincides with $d - N$ for any integer $N$, since then $j_n = -N$ and the $\Gamma ( j_n )$ factor diverges. This is a general feature: while interactions of principal series fields always have a total conformal weight with $\text{Re} \left( \Delta_T \right) > d$ and are free of such divergences, for complementary series fields our adiabatic hypothesis can break down whenever $\text{Re} \left( \Delta_T \right) = d - N$. 
However, that is not to say that every such interaction of light fields leads to problematic divergences in every $S$-matrix element. 
For instance, even though $\sigma^3$ gives a divergent contribution to $\SBD_{0\to 3}$, it gives a finite exchange contribution to $\SBD_{0 \to 4}$ since a non-zero $k_s$ (or $k_t$ or $k_u$) effectively regulates the divergence. Explicitly, we find that for $s$-channel scattering in $d=3$ it is given by
\begin{align}
 \SBD_{0 \to 4}^{\rm exch} &= \frac{\lambda_3^2}{Z_\sigma^6}  \; \frac{  \text{Li}_2 \left( \frac{k_{1} + k_2 - k_s}{k_T} \right) + \text{Li}_2 \left( \frac{k_3 + k_4 - k_s }{ k_T } \right) - \frac{\pi^2}{6}  }{2 i k_s \sqrt{2 i k_1} \sqrt{2 i k_2} \sqrt{2 i k_3} \sqrt{2 i k_4} } \; ,  
 \label{eqn:Sexch_eg}
\end{align}
where we now omit the total-momentum $\delta$-function.

Now consider a massive field $\phi$ coupled to $n-1$ of these $\sigma$ fields, namely $\mathcal{L}_{\rm int} = \sqrt{-g} \frac{ \lambda'_{n} }{ (n-1) !} \sigma^{n-1} \phi$. The $n$-point Bunch-Davies $S$-matrix is 
\begin{align}
\SBD_{0 \to n} = \frac{ i \lambda_{n}'  }{ Z_\sigma^{n-1} Z_\phi }  \frac{  \left| \Gamma \left( \frac{2 j_n + 1 }{2}   - i \mu \right) \right|^2  }{ ( k_\sigma^2 - k_\phi^2 )^{j_n} \sqrt{2 i k_\phi } } \,    \frac{  P^{-j_n}_{i \m - \frac{1}{2}} \left( \frac{k_\sigma}{k_\phi} \right)  }{ \prod_{b=1}^n  \sqrt{2 i k_b} }
\label{eqn:eg_2}
\end{align}
where $k_\sigma \equiv k_1 + ... + k_{n-1}$ is the total energy of the $\sigma$ fields, $k_\phi \equiv k_n$ is the energy of the heavy $\phi$ field and $P^{-j_n}_{i \m - \frac{1}{2}}$ is the associated Legendre polynomial \footnote{Note that since
\begin{align}
P_{0}^{-j} ( z ) = \frac{1}{\Gamma ( 1 + j ) } \left( \frac{z-1}{z + 1 }  \right)^{j/2} \; ,
\end{align}
\eqref{eqn:eg_2} indeed reduces to \eqref{eqn:eg_1} when $i\mu$ is continued to the value $1/2$.
}.

In all of the above examples, when $d=3$ any of the $\sigma$ fields may be replaced by the time derivative of a massless ``pion" field $\dot{\pi}$ by simply multiplying the corresponding $\SBD_{0 \to n}$ by a factor of $i k$ for that external leg.

Note that if we had instead normalised our asymptotic states by a factor of $\sqrt{2 i k}$ per particle, the analytic structure of each of these $S$-matrix elements would be very simple. 
In particular, the only singularities of these $\tilde{\SBD}_{n}$ in the complex $\tilde{k}$-plane would be at $\tilde{k}_T = 0$ when the total energy flowing into the diagram vanishes and, in the case of the exchange diagram \eqref{eqn:Sexch_eg}, also when the energy flowing into either vertex vanishes (namely when $\tilde{k}_L = \tilde{k}_1 + \tilde{k}_2 + k_s$ or $\tilde{k}_R = \tilde{k}_3 + \tilde{k}_4 + k_s$ vanishes). 
Depending on whether each particle is in the initial or final state, these branch points will correspond to the vanishing of a particular linear combination of the $k_b$ (e.g. for $\SBD_{\bfk' \to \bfk_1 \bfk_2 \bfk_3}$, this would be $k_1 + k_2 + k_3 - k' = 0$).

In contrast, the Unruh-DeWitt $S$-matrix elements can have several singularities from each vertex.
For instance, from the $\sigma^n$ interaction considered above,
\begin{align}
\SUD_{1 \to n-1} &= \frac{i \lambda_n}{ Z_\sigma^n \alpha^n  }  \prod_{b=1}^{n-1} \frac{1}{ \sqrt{2 i k_b} } \\
&\times \left[  \frac{\alpha  \Gamma (  j_n  ) }{ ( i k_T - i k' )^{ j_n } \sqrt{-2 i k'} }  -
\frac{ \beta \Gamma (  j_n  ) }{ ( i k_T + i k' )^{ j_n } \sqrt{2 i k'}}  
\right] ,  \nonumber 
\end{align}
where $k'$ is the ingoing momentum and $k_T$ the total outgoing energy.
From the $\sigma^{n-1} \phi$ interaction, 
\begin{align}
\SUD_{1 \to n-1} \propto   \frac{ Q_{i \mu-\frac{1}{2}}^{-j_n} \left(  \frac{k_\sigma}{k_\phi} \right)  }{ ( k_\sigma^2 - k_\phi^2 )^{j_n} \sqrt{ 2 i k_\phi } }  \prod_{b=1}^{n} \frac{1}{ \sqrt{2 i k_b} } \; . 
\end{align}
The former manifestly has singularities at both $k_T = \pm k'$, while the latter has singularities at both $k_\sigma = \pm k_\phi$ (cf. the discussion of $\GUD_2$ above). This illustrates that choosing different bases for the de Sitter $S$-matrix can result in very different singularities. 

As a final example, consider the unique 3-particle $S$-matrix. 
For three general masses, this unique structure can be written explicitly in terms of the Appell $F_4$ function. For example
\begin{align}
    \SUD_{\bfk_1 \bfk_2 \to \bfk_3}  \propto& \,  \frac{1}{   (i k_3 )^{d/2}   } \left( \frac{k_2}{k_3} \right)^{i \m_1} \left( \frac{k_1}{k_3} \right)^{i \m_2}   \\
    &\times F_4 \left( a_+, a_- ; 1 + i \m_1 , 1 + i \m_2; \frac{k_1^2}{k_3^2} , \frac{k_2^2}{k_3^2}  \right) 
    \nonumber
\end{align}
where the indices are $a_{\pm} = \frac{1}{2} \left( \frac{d}{2} + i \m_1 + i \m_2 \pm i \m_3  \right)$. 
For particular mass values (e.g. if the masses coincide or take the special value $i\m = 1/2$) this general Appell function often simplifies into ${}_2 F_1$ functions. 
This example illustrates that the special functions routinely encountered when computing observables in the expanding Poincar\'{e} patch are an unavoidable consequence of these co-ordinates (in particular labelling the fields by $\bfk$) and are not rendered simpler by considering a ``better'' observable such as the $S$-matrix.

\section{Cosmological observables from the $S$-matrix}

\noindent Finally, we turn to the question of how to extract inflationary observables from the de Sitter $S$-matrix. \\

\noindent {\it Wavefunction of the Universe} --- The de Sitter $S$-matrix elements are closely related to non-Gaussianities of the late-time Bunch-Davies wavefunction, which can in fact be constructed from the $S$-matrix once a particular field basis has been chosen. 
Specifically, the wavefunction of a state at time $\tau$ is defined by projecting the state onto a basis of field eigenstates, $| \phi ( \tau ) \rangle$, which are defined by $\hat{\phi} ( \tau, \bfk) | \phi ( \tau ) \rangle = \phi ( \tau, \bfk) | \phi ( \tau ) \rangle$.
The wavefunction of the Bunch-Davies vacuum $| 0 , - \infty \rangle$ is of particular importance in early Universe cosmology since it describes the statistics of primordial perturbations, which seed inhomogeneities in the Cosmic Microwave Background, as well as density perturbations of the Large-Scale Structure of the universe. 
This wavefunction can be characterised by a set of ``wavefunction coefficients,'' which are most conveniently written as the connected part of the following matrix element \footnote{
Here, ``connected'' corresponds to keeping only contributions which are proportional to a single momentum-conserving delta function---perturbatively, this corresponds to keeping only connected Feynman-Witten diagrams. 
} 
\begin{align}
 \psi_n (\tau)  \equiv  \langle \phi ( \tau )  = 0 | \left[ \prod_{b=1}^n \frac{ i \hat{\Pi} ( \tau , \bfk_b ) }{Z^2}  \right] | 0 , -\infty \rangle_{\rm in}  \; . 
 \label{eqn:psi_def}
\end{align}
where $\hat{\Pi}$ is the momentum conjugate to $\hat{\phi}$ \footnote{
This is often written in terms of field eigenstates as
\begin{align}
\langle \phi ( \tau)  \,  | \, 0, -\infty \rangle_{\rm in} = \text{exp} \left(    
\sum_{n}^{\infty} \left[ \prod_{b=1}^n \int \frac{d^d \bfk_b}{(2\pi)^d} \, \phi ( \tau , \bfk_b ) \right]  \, \frac{  \psi_n ( \tau ) }{n!} 
\right) \;  
\end{align}
since in that basis each $i \hat{\Pi}$ operator in \eqref{eqn:psi_def} implements the field derivative $Z^2 \, \delta/\delta \phi$. 
}.

To relate these coefficients to the $S$-matrix elements, we expand the bra in terms of our asymptotic out-states, 
\begin{align}
\lim_{\tau \to 0} \langle \phi  = 0 |   \frac{ i \hat{\Pi}_1 }{Z^2} ... \frac{ i \hat{\Pi}_n}{Z^2}   = \sum_{j}   {}_{\rm out} \langle j , -\infty |  \, B^{j}_{n}   \, , 
\end{align}
since then we can write the wavefunction coefficient as a sum over $0 \to j$ matrix elements, 
\begin{align}
\lim_{\tau \to 0} \psi_n ( \tau ) = 
\sum_j  B^{j}_{n} \;  \SBD_{0 \to j} \; . 
\end{align} 
The $B_n^j$ coefficients can be evaluated in the free theory, since by the adiabatic hypothesis 
\begin{equation}
\begin{split}
| j , -\infty \rangle_{\rm out} \;\; &\to \;\; \hat{a}_1^{\dagger} ... \hat{a}_j^{\dagger }|0 , -\infty \rangle \\ (- \tau )^{d} \hat{\Pi} ( \tau , \bfk )\;\; &\to \;\;  Z^2  \tau \partial_\tau \phi ( \tau, \bfk )
\end{split}
\end{equation}
as $\tau \to 0$ and the interactions turn off.
This gives
\begin{widetext}
\begin{align}
\lim_{\tau \to 0}  \psi_{\bfk_1 ... \bfk_n} ( \tau ) = \sum_{j=0}^{\infty}  
\left[ \prod_{\ell=1}^j \int_{\bfq_{\ell} \bfq_{\ell}'} P_{\bfq_{\ell} \bfq'_{\ell}} (\tau) \right]
\frac{ (-1)^j \SBD_{0 \to \bfk_1 ... \bfk_n \bfq_1 \bfq'_1 ... \bfq_j \bfq'_j } }{
\left[  \prod_{b=1}^n  ( - \tau )^{d/2} f^+ ( k_b \tau ) \right] \left[
 \prod_{c=1}^j  ( - \tau )^{d} f^+ ( q_c \tau ) f^+ ( q_c' \tau )
 \right]
}  \; ,
\label{eqn:psi_from_S}
\end{align}
\end{widetext}
where $P_{\bfq \bfq'} (\tau) = \langle 0 ,-\infty | \hat{\phi} ( \tau, \bfq ) \hat{\phi} ( \tau, \bfq' ) | 0, - \infty \rangle$ is the free-theory power spectrum and $\int_{\bfq \bfq'} \equiv \int \frac{d^d \bfq}{(2 \pi )^d} \frac{d^d \bfq'}{ ( 2\pi )^d}$. 
Equation \eqref{eqn:psi_from_S} allows one to explicitly construct any wavefunction coefficient from the $S$-matrix elements and the mode functions of the fields. 
In practice, this infinite sum will always truncate at a given order in perturbation theory, so often only the first few $S$-matrix elements ($S_{0 \to n}$, $S_{0 \to n+2}$, ...) are needed. 

This is best illustrated with an example. 
Consider the quartic coefficient $\psi_4$ generated by the interactions $\mathcal{L}_{\rm int} = \sqrt{-g} \left( \frac{\lambda_3}{3!} \phi^3 + \frac{\lambda_4}{4!} \phi^4 \right)$ at tree-level.
From the $\phi^4$ interaction, the contact Feynman-Witten diagram gives
\begin{align}
 \psi_{\bfk_1 \bfk_2 \bfk_3 \bfk_4}^{\rm cont} ( \tau' ) &= i \lambda_4 \int_{-\infty}^{\tau'} \frac{d \tau}{- \tau} ( - \tau )^d  \, \prod_{b=1}^4  \, K_{k_b} ( \tau ; \tau'  ) \; 
 \label{eqn:psi4_eg} 
\end{align}
where the ``bulk-to-boundary'' propagator is given in terms of the Hankel mode functions by
\begin{align}
 K_k ( \tau ; \tau' ) = \frac{ ( - \tau )^{d/2} f^+ ( k \tau ) }{ ( - \tau')^{d/2} f^+ ( k \tau' )  } \; .  
\end{align}
Comparing with \eqref{eqn:Sn_cont}, we see that this contribution to the wavefunction can indeed be written as
\begin{align}
 \psi_{\bfk_1 \bfk_2 \bfk_3 \bfk_4}^{\rm cont} ( \tau) &= \frac{ \SBD_{0 \to \bfk_1 \bfk_2 \bfk_3 \bfk_4}^{\rm cont} }{ \prod_{b=1}^4 (- \tau )^{d/2} f^+ ( k_b \tau)  } \; 
\end{align}
at late times (which corresponds to the general formula \eqref{eqn:psi_from_S} with $n=4$ and $j = 0$). 
The exchange contribution $\psi_4^{\rm exch} (\tau)$ is almost identical to \eqref{eqn:S4_exch}, but with $G_{2} ( k_s \tau,  k_s\tau')$ replaced by the ``bulk-to-bulk propagator'',
\begin{align}
&-i G_k^{\rm bulk} (\tau_1, \tau_2 ; \tau )  = G_2 \left( k \tau_1 , k \tau_2 \right) \nonumber \\
 &\qquad\quad\qquad- \frac{ f^- ( k \tau )  }{ f^+ ( k \tau ) }  (- \tau_1 )^{d/2} f^+ ( k \tau_1) ( - \tau_2 )^{d/2} f^+ ( k \tau_2 )  \; . \nonumber
\end{align}
The $\phi^3 \times \phi^3$ contribution to the wavefunction can therefore be written as
\begin{align}
&\psi_{\bfk_1 \bfk_2 \bfk_3  \bfk_4}^{\rm exch} ( \tau ) 
=
\frac{ 1 }{ \prod_{b=1}^4 (- \tau )^{d/2}  f^+ ( k_b \tau )  } \Bigg[ \SBD_{0 \to \bfk_1 \bfk_2 \bfk_3 \bfk_4}^{\rm exch} 
\nonumber \\
&\quad- \int_{\bfq \bfq'}  \frac{  P_{\bfq \bfq'} (\tau) \SBD_{0 \to \bfk_1 \bfk_2 \bfq } \SBD_{0 \to \bfk_3 \bfk_4  \bfq' } }{  (-\tau )^d f^+ ( q \tau) f^+ ( q' \tau ) } - \text{2 perm.} \Bigg] 
\end{align}
at late times. 
So the full $\psi_4^{\rm cont} + \psi_4^{\rm exch}$ is indeed given by \eqref{eqn:psi_from_S}, since the only non-zero $\SBD_{0 \to n}$ at this order in perturbation theory are $\SBD_{0 \to 4} = \SBD^{\rm cont}_{0 \to 4} + \SBD^{\rm exch}_{0 \to 4}$ and the disconnected part of $\SBD_{0 \to 6}$ ($ = \SBD_{0\to 3} \SBD_{0 \to 3}$), which becomes connected once integrated over $\bfq$ and $\bfq'$. 

Note that it is also straightforward to invert \eqref{eqn:psi_from_S} and express a given $S$-matrix element in terms of the wavefunction. For example
\begin{align}
    \SBD_{0 \to 4} = &\lim_{\tau \to 0} \left[ \prod_{b=1}^4 (- \tau)^{\frac{d}{2}} f^+ ( k_b \tau ) \right] \Bigg( 
    \psi_{\bfk_1 \bfk_2 \bfk_3  \bfk_4} ( \tau ) \label{eqn:S4_from_psi} \\
    &+ \int_{\bfq \bfq'} P_{\bfq \bfq'} ( \tau ) \, \psi_{\bfk_1 \bfk_2 \bfq} ( \tau ) \psi_{\bfk_3 \bfk_4 \bfq'} ( \tau) + \text{2 perm.}
    \Bigg) \nonumber 
\end{align}
Previous results for the wavefunction coefficients can then be readily translated into $S$-matrix elements. \\

\noindent {\it In-in correlators} ---
We can similarly extract from the $S$-matrix any desired equal-time correlator at late times.
These objects are the closest to what one would observe in primordial non-Gaussianity, and are defined by \footnote{
The correlator can also be defined in terms of the wavefunction coefficient using the functional integral
\begin{align}
\langle \phi_1 ... \phi_n \rangle   = \frac{ \int \mathcal{D} \phi \; \phi_1 ... \phi_n  | \langle \phi | 0 , - \infty \rangle_{\rm in} |^2  }{ \int \mathcal{D} \phi \; | \langle \phi | 0 , - \infty \rangle_{\rm in} |^2 } \; . 
\end{align}
\\[-12pt] \!
}
\begin{align}
\langle \phi_1 ... \phi_n \rangle  \equiv \lim_{\tau \to 0}  {}_{\rm in} \langle 0  | \, \hat{\phi} ( \tau, \bfk_1 ) ... \hat{\phi} (\tau, \bfk_n) | 0  \rangle_{\rm in}  \; .
\end{align}
If we therefore decompose the product
\begin{align}
\lim_{\tau \to 0} \hat{\phi}_1  ... \hat{\phi}_n  = {\textstyle \sum_{j, j'}}   |j , - \infty \rangle  {}_{\rm out}  \; C_n^{j j'}  \;   {}_{\rm out} \langle j' , - \infty | 
\label{eqn:phi_prod_exp}
\end{align}
and use the definition of the $S$-matrix \eqref{eqn:Sdef2}, we have that
\begin{align}
\langle \phi_1 ... \phi_n \rangle  =  {\textstyle \sum_{j, j'}}  \; \SBD_{0 \to j}^* \; C_n^{j j'} \; \SBD_{0 \to j'} \; . 
\end{align} 
Since the field operators and out-states in \eqref{eqn:phi_prod_exp} coincide with those of the free theory at $\tau \to 0$, we can immediately evaluate the $C_n^{j j'}$ 
in terms of the late-time limit of $(-\tau)^{d/2} f^{\pm} ( k \tau)$, which we denote by $f_k^{\pm}$. This produces a very explicit relation between the observable correlator and the de Sitter $S$-matrix:
\begin{widetext}
\begin{align}
\langle \phi_1 ... \phi_n \rangle  =   2 \text{Re} \left[ 
 \sum_{j'=0}^{n/2} 
\left[ \prod_{b=1}^{n-j'}  f^-_{k_b} \right] \left[ \prod_{b=n-j'+1}^{n}  f^+_{ k_b }  \right]  
 \sum_{j=0}^{\infty}   
  \int_{\substack{ \bfq_1 ... \bfq_j \\ \bfq_1' ... \bfq_j' }} 
   \SBD_{0 \to \bfk_1 ... \bfk_{n-j'} \bfq_1 ... \bfq_{j} } 
   \SBD_{0 \to \bfk_{n-j'+1} ... \bfk_n \bfq'_1 ... \bfq'_{j} }^*  + \text{perm.}
   \right]
\label{eqn:corr_from_S}
\end{align}
\end{widetext}
where $\int_{ \substack{ \bfq_1 ... \bfq_j \\ \bfq_1' ... \bfq_j' } }$ is an integral over the pairs $( \bfq_{\ell} , \bfq_{\ell}')$ subject to the condition $\bfq_{\ell} + \bfq_{\ell}' = 0$. The ``$+ \text{perm.}$'' denotes a symmetrisation over all possible permutations of the external $\bfk$'s. 
We can immediately notice some differences with the wavefunction relation \eqref{eqn:psi_from_S}: the correlators
\begin{itemize}

\item depend quadratically on the $S$-matrix elements, while the wavefunction coefficients are linear, 

\item are manifestly real, since $\phi$ is real Hermitian and we have assumed spatial parity (so $\langle \phi_{\bfk_1} ... \phi_{\bfk_n} \rangle = \langle \phi_{-\bfk_1} ... \phi_{-\bfk_n} \rangle $),  

\item are insensitive to phase information in the $S$-matrix (which would determine the late-time mixed correlators of $\hat{\phi}$ and its conjugate momentum $\hat{\Pi}$).  

\end{itemize}

As a concrete example, consider the four-point correlator induced by the same Lagrangian $\mathcal{L}_{\rm int} = \sqrt{-g} \left( \frac{\lambda_3}{3!} \phi^3 + \frac{\lambda_4}{4!} \phi^4 \right)$. 
This correlator can be determined from the wavefunction coefficients studied above, and can be separated into three different contributions, 
\begin{align}
&\langle \phi_1 ... \phi_4 \rangle^{\rm cont} =  2 \text{Re} \left[ \psi_4^{\rm cont} \right] {\textstyle \prod_{b=1}^4 | f_{k_b}^+ |^2 } \; ,  \nonumber \\
&\langle \phi_1 ... \phi_4 \rangle^{\rm exch} =  2 \text{Re} \left[ \psi_4^{\rm exch} \right]  {\textstyle \prod_{b=1}^4 | f_{k_b}^+ |^2  } \; ,   \label{eqn:corr_from_psi}  \\
&\langle \phi_1 ... \phi_4 \rangle^{\rm quad} = {\textstyle \prod_{b=1}^4 | f_{k_b}^+ |^2 } \nonumber \\
&\times \left( \int_{\bfq \bfq'} P_{\bfq \bfq'} \, 2 \text{Re} \left[ \psi_{\bfk_1 \bfk_2 \bfq } \right]  2 \text{Re} \left[ \psi_{\bfk_3 \bfk_4 \bfq' } \right] + \text{2 perm.} \right)   . \nonumber
\end{align}
Let us now confirm that this is correctly reproduced by \eqref{eqn:corr_from_S}. 
At this order in perturbation theory, there are only three $S$-matrix elements that contribute, namely $\SBD_{0 \to 3}$, $\SBD_{0 \to 4}$ and $\SBD_{0 \to 6} \supset \SBD_{0 \to 3} \SBD_{0 \to 3}$.
Comparing their integral representations with \eqref{eqn:corr_from_psi}, we see that
\begin{align}
&\langle \phi_{1}  ... \phi_{4} \rangle^{\rm cont} =  2 \text{Re} \left[  {\textstyle \prod_{b=1}^4 } f_{k_b}^- \; \SBD^{\rm cont}_{0 \to \bfk_1 ... \bfk_4} \right]    
\end{align}
\begin{align}
&\langle \phi_{1}  ... \phi_{4} \rangle^{\rm exch} =  2 \text{Re} \Bigg[ {\textstyle \prod_{b=1}^4 } f_{k_b}^-   \nonumber \\
&\qquad\qquad \times \left( \SBD^{\rm exch}_{0 \to \bfk_1 ... \bfk_4} 
- \int_{\bfq \bfq'} \frac{ P_{\bfq \bfq'}  }{ f_q^+ f_{q'}^+ } \SBD_{0 \to \bfk_1 ...\bfk_4 \bfq \bfq'}     \right)       \Bigg]   \nonumber  \\ \nonumber 
&\langle \phi_{1}  ... \phi_{4} \rangle^{\rm quad} =   
2 \text{Re} \Bigg[ 
 \int_{\bfq \bfq'} \frac{ P_{\bfq \bfq'} }{f^+_{q} f^+_{q'}} \SBD_{0 \to \bfk_1 ... \bfk_4 \bfq \bfq'} {\textstyle \prod_{b=1}^4 }  f_{k_b}^-   \nonumber \\
&\qquad+ f_{k_1}^- f_{k_2}^- f_{k_3}^+ f_{k_4}^+  \int_{\bfq \bfq'} \SBD_{0 \to \bfk_1 \bfk_2 \bfq} \SBD^*_{0 \to \bfk_3 \bfk_4 \bfq'}  + \text{perm.} 
\Bigg]
\nonumber 
\end{align}
and so the sum of all three contributions to the correlator is indeed given by the general relation \eqref{eqn:corr_from_S}.

\section{Future directions}

\noindent In summary, we have defined a perturbative $S$-matrix for scalar fields in the expanding patch of a fixed de Sitter spacetime background, and demonstrated that it enjoys many of the useful properties of the Minkowski $S$-matrix. 
We believe that fundamental properties like unitarity, causality and locality will be simpler to express in terms of these $S$-matrix elements (as opposed to, say, the wavefunction or in-in correlators). This expectation stems from the fact that this $S$-matrix describes the time evolution in a field-independent way, and is the natural extrapolation of the Minkowski $S$-matrix to non-zero values of the Hubble rate.
They are also in many cases easier to compute and analyse than their wavefunction counterparts. In an upcoming companion paper \cite{us}, we will describe in more detail how to efficiently compute these $S$-matrix elements.   \\

 \noindent {\it Unitarity} --- Interestingly, the particular combination of wavefunction coefficients that corresponds to an $S$-matrix element was previously constructed in \cite{Cespedes:2020xqq} via an independent argument. There, this combination (which we shall denote at finite times by $\tilde{\psi}_n (\tau)$), was engineered as the unique combination of the wavefunction coefficients which remains invariant under the free evolution for any initial condition. Constraints from unitarity were therefore formulated most simply in terms of $\tilde{\psi}_n (\tau)$ because its time dependence stems only from the interactions, whereas the time-dependence of the original $\psi_n (\tau)$ is a convolution of both the interactions and the initial condition. Here we have uncovered a deeper reason for the simplicity of $\tilde{\psi_n} ( \tau)$: it is a finite-time counterpart to the $S$-matrix (and coincides with the $S$-matrix as $\tau \to 0$).  
 Furthermore, recent cutting rules for wavefunction coefficients have found a proliferation of terms not present in the usual Cutkosky rules on Minkowski due to the presence of the boundary term in $G^{\rm bulk}_k$ \cite{Goodhew:2020hob, Cespedes:2020xqq, Melville:2021lst, Goodhew:2021oqg, Baumann:2021fxj, Albayrak:2023hie, Agui-Salcedo:2023wlq}. Since the de Sitter $S$-matrix uses the Feynman propagator for internal lines, it obeys simpler cutting rules than the wavefunction of the universe. They are essentially identical to the usual Cutkosky rules. Finally, the analytic continuation to negative values of $| \bfk |$ which has played a central role in cosmological cutting rules can now be understood as a crossing transformation which relates the $0 \to n$ matrix element to a conjugate channel such as $n \to 0$, which therefore recovers the usual form of the optical theorem.  \\

 \noindent {\it Analyticity} --- Although we have focussed on scattering particles with an on-shell energy $\tilde{k} = \pm | \bfk|$, our definition of $\tilde{\SBD}_n$ in \eqref{eqn:Stilde_def} can be evaluated at any value of $\tilde{k}$. In particular, since $f^+ ( \tilde{k} \tau ) \sim e^{+ i \tilde{k} \tau }$ and the integration domain restricts $\tau < 0$, the off-shell $\tilde{\SBD}_n$ must be \emph{analytic} in the upper half of the complex $\tilde{k}$-plane (for any time-ordered correlator that is exponentially bounded at large $\tau$). This is the precise analogue of the Kramers-Kronig analyticity that underpins non-relativistic dispersion relations. $\tilde{S}_n$ is therefore a natural extension of the off-shell wavefunction of \cite{Salcedo:2022aal} to de Sitter spacetime.  \\

 \noindent {\it Locality} --- One crucial consequence of locality (together with unitarity and analyticity) on Minkowski is the ``Froissart bound,'' which limits the growth of scattering amplitudes at large centre-of-mass energies. Since our $S$-matrix reduces to the Minkowski $S$-matrix in the high-energy limit $k \to \infty$, we expect that a similar bound will apply to the growth of the de Sitter $S$-matrix. A rigorous proof of this is left for the future.  \\

 \noindent {\it Renormalisation} --- The divergences which appear at the conformal boundary as $\tau \to 0$ and complicate the $S$-matrix for light fields are similar to the ones encountered near the conformal boundary of AdS. For the latter, there is a well-understood procedure of holographic renormalisation. Perhaps in most cases the $S$-matrix for light fields can be safely defined (or at least reliably computed in perturbation theory) by applying an analogous renormalisation procedure on de Sitter. Progress in that direction would extend the $S$-matrix construction described here to fields of any stable mass (in principal or complementary series).

\noindent {\it Other off-shell extensions} --- Finally, we note that the extension $k \to \tilde{k}$ is not the only way to define an ``off-shell'' $S$-matrix. In particular, another option is to replace the mass parameter $\m \to \tilde{\m}$ (which is now independent of the mass $m^2$), and interpret the LSZ reduction formula as a Kontorovich-Lebedev integral transform from $\tau$ to $\tilde{\mu}$. This alternative procedure for going off-shell has a closer connection to the K\"all\'en-Lehmann spectral representation of flat space, and we aim to discuss it further in \cite{us}. \\

 \noindent {\it Acknowledgements} --- It is a pleasure to thank Dionysios Anninos, Santiago Ag\"{u}\'{i} Salcedo, Tarek Anous, Nima Arkani-Hamed, Daniel Baumann, Paolo Benincasa, Carlos Duaso Pueyo, Austin Joyce, Hayden Lee, Juan Maldacena, Enrico Pajer, Sasha Polyakov, and Dong-Gang Wang for many enlightening discussions on de Sitter space over the years. SM is supported by a UKRI Stephen Hawking Fellowship (EP/T017481/1). GLP is supported by Scuola Normale, by a Rita-Levi Montalcini fellowship from the Italian Ministry of Universities and Research (MUR), and by INFN (IS GSS-Pi).

\section{Appendix}
\appendix
\section{Adiabatic hypothesis}

\noindent Here we give a technical account of our ``adiabatic hypothesis,'' which is the assumption that the interactions turn off sufficiently quickly in the far past/future so that asymptotic states in the interacting theory are reliably captured by the corresponding states in the free theory. 
In particular, we wish to highlight that while adiabaticity in the far past ($\tau \to -\infty$) follows from essentially the same argument as in Minkowski space, the far future of de Sitter ($\tau \to 0$) is qualitatively different. In particular, the adiabatic hypothesis only strictly applies for sufficiently massive fields. 
We will also focus on modes with $\bfk \neq 0$ which regulates possible IR divergences.

The basic idea is that $\hat{\varphi} ( \tau, \bfk )$ acting on the vacuum should produce a new state which contains (with some non-zero probability) a single particle of momentum $\bfk$. 
We write this probability amplitude as
\begin{align}
 {}_{\rm out} \langle \bfq , - \infty | \hat{\varphi} ( \tau, \bfk ) | 0 , - \infty \rangle_{\rm out} \equiv f_{\rm out}^+ ( k \tau ) \, (2 \pi )^d \delta^d \left( \bfk + \bfq \right) 
\end{align}
for the out-states, and
\begin{align}
 {}_{\rm in} \langle \bfq , - \infty | \hat{\varphi} ( \tau, \bfk ) | 0 , - \infty \rangle_{\rm in} \equiv f_{\rm in}^+ ( k \tau ) \, (2 \pi )^d \delta^d \left( \bfk + \bfq \right) 
\end{align}
for the in-states.
If $\hat{\varphi}$ is an operator in the principal series, then $f_{\rm out}^+$ and $f_{\rm in}^+$ take the same form as $f^+$ in \eqref{eqn:f1_def} but with a possibly renormalized $Z$ and $\mu$. 

The technical issue is that, in the interacting theory, $\hat{\varphi}$ can also create multiple particles.
In particular, while $Z^2 f^-  ( \overset{\leftrightarrow}{ \tau \partial_\tau} ) \hat{\varphi}$ creates a normalised one-particle state in the free theory, in the interacting theory it creates:
\begin{align}
 &i Z_{\rm out}^2 f_{\rm out}^-  ( \overset{\leftrightarrow}{ \tau \partial_\tau} ) \hat{\varphi} ( \tau, \bfk ) | 0 , - \infty \rangle_{\rm out}  \nonumber \\
&= 
  |\bfk , - \infty \rangle_{\rm out} + \sum_{n=2}^{\infty} c_{\rm out} ( k \tau  ; n ) | n , -\infty \rangle_{\rm out}
\end{align}
where the $c_{\rm out}$ are the probability amplitudes for creating $n$ particles from the vacuum. 
There is an analogous set of $c_{\rm in}$ defined by $Z_{\rm in}^2 f_{\rm in}^-  ( \overset{\leftrightarrow}{ \tau \partial_\tau} ) \hat{\varphi}$ acting on the in-vacuum
The adiabatic hypothesis that interactions `switch off' at early and late times is formally the requirement that
\begin{align}
\lim_{\tau \to 0} \, c_{\rm out} ( k \tau ; n )  &= 0 \; , 
&\lim_{\tau \to - \infty} \, c_{\rm in} ( k \tau ; n )  &= 0 \; .
\end{align}

The vanishing of $c_{\rm in}$ at early times is ultimately the same assumption that is made to define the Minkowski $S$-matrix, and since de Sitter is indistinguishable from Minkowski as $k \tau \to -\infty$ the usual arguments can be used to justify this weak limit \cite{Lehmann:1954rq} (although see \cite{Lehmann:1954rq, Haag:1958vt, Haag:1959ozr, ruelle1962asymptotic, cmp/1103758732} and more recently \cite{Collins:2019ozc} for subtleties related to composite or unstable particles). 
The only qualitative difference is that any finite mass parameter $\m$ will blue-shift away in the far past, and all such fields behave essentially as if massless---this can lead to the same kind of IR divergences which appear for massless particles on Minkowski, but these do not affect the $S$-matrix for scattering hard modes with $k \neq 0$.

The vanishing of $c_{\rm out}$ at late times is more subtle, and our argument for this is essentially perturbative.
Since in the Heisenberg picture the time evolution of $\hat{\varphi}$ is determined by the equation of motion 
\begin{align}
   Z^2  \mathcal{E} [ k \tau ] \varphi ( \tau, \bfk ) =  \tau \, \frac{\delta S_{\rm int}}{\delta \varphi ( \tau, \bfk) } \,,
\end{align}
we can write the general solution as
\begin{align}
\varphi (\tau, \bfk) =& f_{\rm out}^+(k \tau)  a_{\rm out}^\dagger ( \bfk) + f_{\rm out}^- ( k \tau ) a_{\rm out} ( -\bfk) \nonumber \\
&+ \int_\tau^0 d \tau' \, G_{\rm out}^{\rm ret} ( k \tau, k \tau' ) \, \frac{\delta S_{\rm int} }{ \delta \varphi ( \tau', \bfk ) } \, ,
\end{align}
where $S_{\rm int}$ is the (suitably renormalized) non-linear part of the action, $\hat{a}_{\rm out}$ annihilates $| 0 , - \infty \rangle_{\rm out}$, and $G_{\rm out}^{\rm ret}$ is the retarded propagator built from $f_{\rm out}$ mode functions. 
One can then verify that de Sitter invariant interactions for massive fields in $S_{\rm int}$ will give contributions to $c_{\rm out}$ that vanish at late times.  
For example, the interaction $S_{\rm int} = \lambda \sqrt{-g} \phi^3$ appears at first order in $\lambda$ as the coefficient
\begin{align}
 &c_{\rm out} ( k \tau; q_1 \tau, q_2 \tau ) \nonumber \\
 &= \lambda \int_{\tau}^0 \frac{d \tau'}{\tau'} ( - \tau' )^{\frac{d}{2}}  \, f_{\rm out}^- ( k \tau' ) f_{\rm out}^+ ( q_1 \tau' ) f_{\rm out}^+ ( q_2 \tau' ) \, , 
\end{align}
which describes the overlap with the 2-particle state $|\bfq_1 \bfq_2 , - \infty \rangle_{\rm out}$ (and we have suppressed a factor of $\delta^d ( \bfq_1 + \bfq_2 - \bfk )$). 
For fields in the principal series, this integral is finite for all $\tau$ and vanishes as $\tau \to 0$.

In general, for an $n$-point interaction involving both light and heavy fields, the integrand which appears in $c_{\rm out}$ behaves like $\sim \tau^{\alpha - 1}$ at small $\tau$, with 
\begin{align}
    \alpha = \frac{d}{2} ( n - 2 ) - \sum_{b=1}^n | \text{Im}\, \mu_b |
\end{align} 
The corresponding integral therefore diverges if the total $| \text{Im} \,\mu_T |$ exceeds $\frac{d}{2} ( n-2 )$, and in that case $c_{\rm out}$ is no longer guaranteed to vanish at late times \footnote{
Though note that even when $\alpha < 0$ the integral may still converge and $c_{\rm out}$ vanish. An example of this would be the $\tilde{\SBD}_3$ from the cubic interaction $\sigma^3$ given in \eqref{eqn:eg_1}: this particular integral is perfectly finite $d=2$ ($\alpha = -1/2$). So while $\alpha > 0$ is sufficient for the adiabatic hypothesis, it is not necessary for particular $S$-matrix elements.
}.

However, there are nonetheless some interactions involving light fields which switch off sufficiently fast to avoid any issues at late times. 
An example would be the cubic interaction $\phi^2 \pi$ between two heavy fields $\phi$ and a light field $\pi$, for which $c_{\rm out}$ vanishes for every non-zero $\pi$ mass. 
More generally, if the late-time divergence in $c_{\rm out}$ is suitably regularised, a renormalized theory of the boundary degrees of freedom may have well-defined $S$-matrix elements for an even wider range of interactions and mass values. 
We leave a systematic exploration of this for the future.

\begin{figure*}[htpb!]
\qquad\qquad\qquad ~ 
$\hphantom{\alpha^2}\SBD^{\rm free}_{2 \to 2} = $
	\begin{tikzpicture}[scale=0.6,baseline=0.85cm]
	\begin{feynman}
				
		\vertex (a) at (0,0) {$ $};
		\vertex (b) at ( 0, 3) {$ $};
		\vertex (c) at (1,0) {$ $};
		\vertex (d) at ( 1, 3) {$ $};
		\diagram* {
			(a) -- (b),
			(c) -- (d)
		};
		
	\end{feynman}
\end{tikzpicture} 
$+$
	\begin{tikzpicture}[scale=0.6,baseline=0.85cm]
	\begin{feynman}
				
		\vertex (a) at (0,0) {$ $};
		\vertex (b) at ( 0, 3) {$ $};
		\vertex (c) at (2,0) {$ $};
		\vertex (d) at ( 2, 3) {$ $};

            \diagram* {
			(a) -- (d)
		};

		\node [thick, draw=white, fill=white, rotate=-30, rectangle, minimum width=0.2cm, minimum height=0.25cm,align=center] at (1,1.5);

        \diagram* {
			(b) -- (c)
		};

	\end{feynman}
\end{tikzpicture} 
\;\; , \hfill
$ \alpha^2 \SUD^{\rm free}_{2 \to 2} = $
	\begin{tikzpicture}[scale=0.6,baseline=0.85cm]
	\begin{feynman}
				
		\vertex (a) at (0,0) {$  $};
		\vertex (b) at ( 0, 3) {$ $};
		\vertex (c) at (1,0) {$ $};
		\vertex (d) at ( 1, 3) {$ $};
		\diagram* {
			(a) -- (b),
			(c) -- (d)
		};
		
	\end{feynman}
\end{tikzpicture} 
$ + $
	\begin{tikzpicture}[scale=0.6,baseline=0.85cm]
	\begin{feynman}
				
		\vertex (a) at (0,0) {$ $};
		\vertex (b) at ( 0, 3) {$  $};
		\vertex (c) at (2,0) {$  $};
		\vertex (d) at ( 2, 3) {$ $};
		\diagram* {
			(a) -- (d)
		};

		\node [thick, draw=white, fill=white, rotate=-30, rectangle, minimum width=0.2cm, minimum height=0.25cm,align=center] at (1,1.5);

        \diagram* {
			(b) -- (c)
		};

	\end{feynman}
\end{tikzpicture} 
 $+  | \beta |^2$
	\begin{tikzpicture}[scale=0.6,baseline=0.85cm]
	\begin{feynman}
				
		\vertex (a) at (0,0) {$ $};
		\vertex (b) at ( 0, 3) {$  $};
		\vertex (c) at (2,0) {$ $};
		\vertex (d) at ( 2, 3) {$  $};
		\diagram* {
			(a) -- [half left] (c),
			(b) -- [half right] (d)
		};
		
	\end{feynman}
\end{tikzpicture}
\qquad\qquad\qquad ~ 
\caption{The free theory $S$-matrix element for ``$2 \to 2$ scattering'' in both the Bunch-Davies and Unruh-DeWitt basis. The latter contains an additional contribution from particle production. A line joining two external points represents a momentum conserving $\delta$ function. 
}
\label{fig:free}
\end{figure*}

\begin{figure*}[htbp!]
\raggedright
$\hphantom{\alpha^4}\SBD_{2 \to 2} = \SBD_{2 \to 2}^{\rm free} \; + $
	\begin{tikzpicture}[scale=0.6,baseline=0.85cm]
	\begin{feynman}
		
		\vertex (C1) at (0.8, 1.5);
		\vertex (C2) at (1.2, 1.5);		
		
		\vertex (a) at (0,0) {$f^- $};
		\vertex (b) at ( 0, 3) {$ f^+ $};
		\vertex (c) at (2,0) {$\;\; f^- $};
		\vertex (d) at ( 2, 3) {$\;\; f^+ $};
		\diagram* {
			(a) --  (C1),
			(b) -- (C1),
			(c) -- (C2),
			(d) -- (C2)
		};
		
		\node [thick, draw=black, fill=black!10, ellipse, minimum width=0.75cm,minimum height=0.45cm,align=center] at (1,1.5);

	\node at (1.05,1.5) {$ G_4 $};				
	
	\end{feynman}
\end{tikzpicture}
$+$
	\begin{tikzpicture}[scale=0.6,baseline=0.85cm]
	\begin{feynman}
				
		\vertex (a) at (0,0) {$f^- $};
		\vertex (b) at ( 0, 3) {$f^+$};
		\vertex (c) at (1,0) {$ $};
		\vertex (d) at ( 1, 3) {$ $};
		\diagram* {
			(a) -- (b),
			(c) -- (d)
		};
		
		\node [thick, draw=black, fill=black!10, ellipse, minimum width=0.45cm,minimum height=0.75cm,align=center] at (0,1.5);
				
	\node at (0,1.5) {$ G_2 $};
	\end{feynman}
\end{tikzpicture} 
$ + $
	\begin{tikzpicture}[scale=0.6,baseline=0.85cm]
	\begin{feynman}
	
		\vertex (a) at (0,0) {$ $};
		\vertex (b) at ( 0, 3) {$ $};
		\vertex (c) at (1,0) {$ \;\; f^- $};
		\vertex (d) at ( 1, 3) {$ \;\; f^+ $};
		\diagram* {
			(a) -- (b),
			(c) -- (d)
		};
		
		\node [thick, draw=black, fill=black!10, ellipse, minimum width=0.45cm,minimum height=0.75cm,align=center] at (1,1.5);
				
	\node at (1,1.5) {$ G_2 $};

	\end{feynman}
\end{tikzpicture} 
$+$
	\begin{tikzpicture}[scale=0.6,baseline=0.85cm]
	\begin{feynman}
				
		\vertex (a) at (0,0) {$ $};
		\vertex (b) at ( 0, 3) {$ f^+ $};
		\vertex (c) at (2,0) {$ \;\; f^- $};
		\vertex (d) at ( 2, 3) {$ $};
		\diagram* {
			(a) -- (d),
			(b) -- (c)
		};

		\node [thick, draw=white, fill=white, rotate=-30, rectangle, minimum width=0.2cm, minimum height=0.75cm,align=center] at (1,1.5);

		\node [thick, draw=black, rotate = 30, fill=black!10, ellipse, minimum width=0.45cm,minimum height=0.75cm,align=center] at (1,1.5);

	\node at (1,1.5) {$ G_2 $};
						
	\end{feynman}
\end{tikzpicture} 
$+$
	\begin{tikzpicture}[scale=0.6,baseline=0.85cm]
	\begin{feynman}
				
		\vertex (a) at (0,0) {$f^- $};
		\vertex (b) at ( 0, 3) {$ $};
		\vertex (c) at (2,0) {$ $};
		\vertex (d) at ( 2, 3) {$ f^+  $};
		\diagram* {
			(a) -- (d),
			(b) -- (c)
		};

		\node [thick, draw=white, fill=white, rotate=+30, rectangle, minimum width=0.2cm, minimum height=0.75cm,align=center] at (1,1.5);

		\node [thick, draw=black, rotate = -30, fill=black!10, ellipse, minimum width=0.45cm,minimum height=0.75cm,align=center] at (1,1.5);

	\node at (1,1.5) {$ G_2 $};
						
	\end{feynman}
\end{tikzpicture} 
\\
\centerline{
$\alpha^4 \SUD_{2 \to 2} = \alpha^4 \SUD_{2\to 2}^{\rm free} + $ 
	\begin{tikzpicture}[scale=0.6,baseline=0.85cm]
	\begin{feynman}
		
		\vertex (C1) at (0.8, 1.5);
		\vertex (C2) at (1.2, 1.5);		
		
		\vertex (a) at (0,0) {$\fUD^- $};
		\vertex (b) at ( 0, 3) {$ f^+ $};
		\vertex (c) at (2,0) {$\;\; \fUD^- $};
		\vertex (d) at ( 2, 3) {$\;\; f^+ $};
		\diagram* {
			(a) --  (C1),
			(b) -- (C1),
			(c) -- (C2),
			(d) -- (C2)
		};
		
		\node [thick, draw=black, fill=black!10, ellipse, minimum width=0.75cm,minimum height=0.45cm,align=center] at (1,1.5);

	\node at (1.05,1.5) {$ \GUD_4 $};				
	
	\end{feynman}
\end{tikzpicture}
$+ \, \alpha \left(\rule{0cm}{1.6cm}\right.$
	\begin{tikzpicture}[scale=0.6,baseline=0.85cm]
	\begin{feynman}
				
		\vertex (a) at (0,0) {$\fUD^- $};
		\vertex (b) at ( 0, 3) {$f^+$};
		\vertex (c) at (1,0) {$ $};
		\vertex (d) at ( 1, 3) {$ $};
		\diagram* {
			(a) -- (b),
			(c) -- (d)
		};
		
		\node [thick, draw=black, fill=black!10, ellipse, minimum width=0.45cm,minimum height=0.75cm,align=center] at (0,1.5);
				
	\node at (0,1.5) {$ \GUD_2 $};
	\end{feynman}
\end{tikzpicture} 
$ + $
	\begin{tikzpicture}[scale=0.6,baseline=0.85cm]
	\begin{feynman}
	
		\vertex (a) at (0,0) {$ $};
		\vertex (b) at ( 0, 3) {$ $};
		\vertex (c) at (1,0) {$ \;\; \fUD^- $};
		\vertex (d) at ( 1, 3) {$ \;\; f^+ $};
		\diagram* {
			(a) -- (b),
			(c) -- (d)
		};
		
		\node [thick, draw=black, fill=black!10, ellipse, minimum width=0.45cm,minimum height=0.75cm,align=center] at (1,1.5);
				
	\node at (1,1.5) {$ \GUD_2 $};

	\end{feynman}
\end{tikzpicture} 
$+$
	\begin{tikzpicture}[scale=0.6,baseline=0.85cm]
	\begin{feynman}
				
		\vertex (a) at (0,0) {$ $};
		\vertex (b) at ( 0, 3) {$ f^+ $};
		\vertex (c) at (2,0) {$ \;\; \fUD^- $};
		\vertex (d) at ( 2, 3) {$ $};
		\diagram* {
			(a) -- (d),
			(b) -- (c)
		};

		\node [thick, draw=white, fill=white, rotate=-30, rectangle, minimum width=0.2cm, minimum height=0.75cm,align=center] at (1,1.5);

		\node [thick, draw=black, rotate = 30, fill=black!10, ellipse, minimum width=0.45cm,minimum height=0.75cm,align=center] at (1,1.5);

	\node at (1,1.5) {$ \GUD_2 $};
						
	\end{feynman}
\end{tikzpicture} 
$+$
	\begin{tikzpicture}[scale=0.6,baseline=0.85cm]
	\begin{feynman}
				
		\vertex (a) at (0,0) {$\fUD^- $};
		\vertex (b) at ( 0, 3) {$ $};
		\vertex (c) at (2,0) {$ $};
		\vertex (d) at ( 2, 3) {$ f^+  $};
		\diagram* {
			(a) -- (d),
			(b) -- (c)
		};

		\node [thick, draw=white, fill=white, rotate=+30, rectangle, minimum width=0.2cm, minimum height=0.75cm,align=center] at (1,1.5);

		\node [thick, draw=black, rotate = -30, fill=black!10, ellipse, minimum width=0.45cm,minimum height=0.75cm,align=center] at (1,1.5);

	\node at (1,1.5) {$ \GUD_2 $};
						
	\end{feynman}
\end{tikzpicture} 
$+  \beta^*$
	\begin{tikzpicture}[scale=0.6,baseline=0.85cm]
	\begin{feynman}
				
		\vertex (a) at (0,0) {$ $};
		\vertex (b) at ( 0, 3) {$ f^+ $};
		\vertex (c) at (2,0) {$ $};
		\vertex (d) at ( 2, 3) {$\;\; f^+ $};
		\diagram* {
			(a) -- [half left] (c),
			(b) -- [half right] (d)
		};
		
		\node [thick, draw=black, fill=black!10, ellipse, minimum width=0.75cm,minimum height=0.45cm,align=center] at (1,1.75);
				
	\node at (1.05,1.75) {$ \GUD_2 $};
	\end{feynman}
\end{tikzpicture}
$+ \beta$ \begin{tikzpicture}[scale=0.6,baseline=0.85cm]
	\begin{feynman}
				
		\vertex (a) at (0,0) {$ \fUD^- $};
		\vertex (b) at ( 0, 3) {$ $};
		\vertex (c) at (2,0) {$ \;\; \fUD^- $};
		\vertex (d) at ( 2, 3) {$ $};
		\diagram* {
			(a) -- [half left] (c),
			(b) -- [half right] (d)
		};
		
		\node [thick, draw=black, fill=black!10, ellipse, minimum width=0.75cm,minimum height=0.45cm,align=center] at (1,1.25);
	\node at (1.05,1.25) {$ \GUD_2 $};

	\end{feynman}
\end{tikzpicture} $ \left. \rule{0cm}{1.6cm}\right)$
}
\caption{In the interacting theory, the $S$-matrix element for ``$2 \to 2$ scattering'' is given by the free contribution shown in Figure~\ref{fig:free} plus the diagrams shown above. 
The Unruh-DeWitt basis again contains additional disconnected diagrams due to the free theory particle production. A line joining two external points represents a momentum-conserving $\delta$ function, and the grey blobs represent the amputated Greens functions shown (which are then put on-shell using the mode functions shown).}
\label{fig:int}
\end{figure*}

\section{Disconnected components}

\noindent In the main text we focused on the connected contributions to the $S$-matrix.
The complete LSZ formula also contains disconnected contributions. 
These arise from the non-zero commutators between $\hat{a}_{\bfk}$, $\hat{a}_{\bfk}^\dagger$, $\hat{b}_{\bfk}$ and $\hat{b}^{\dagger}_{\bfk}$, where the $\hat{b}$ operators are the Unruh-DeWitt analogue of the $\hat{a}$ operators, i.e. they create $|1, 0 \rangle$ from $|0,0 \rangle$ in the free theory. 
Specifically, the LSZ derivation in \eqref{eqn:LSZ_derivation} also produces disconnected terms such as
\begin{align}
    \hat{a}_{\bfk} | n \rangle_{\rm in} &= \sum_{b=1}^n (2 \pi)^d \delta^d \left( \bfk - \bfk_b \right) | n - 1 \rangle_{\rm in} \; ,   \label{eqn:disc_comp} \\
   \alpha | n \rangle_{\rm in} &=  \hat{b}_{\bfk_n}^\dagger | n - 1 \rangle_{\rm in} - \beta \sum_{b=1}^{n-1} (2 \pi)^d \delta^d \left( \bfk_n - \bfk_b \right) | n - 1 \rangle_{\rm in} \, . \nonumber 
\end{align}
For instance, the full expression for the $2 \to 2$ $S$-matrix elements are shown diagrammatically in Figures \ref{fig:free} and \ref{fig:int}.

\bibliographystyle{apsrev4-1}
\bibliography{dS_S_Matrix.bib}

\end{document}